%% file: draft_clean.tex
\documentclass[fleqn,usenatbib]{mnras}

\usepackage{newtxtext,newtxmath}

\usepackage{fixltx2e} 
\usepackage{graphicx, amsmath, amssymb, aas_macros, natbib, tabularx, ulem}
\usepackage{bm}
\input{extra_preamble.tex}

\newcommand{\xieff}{\xi_{\rm eff}}
\newcommand{\fsurv}{f_{\rm surv}}
\newcommand{\fescobs}{f_{1500}}
\newcommand{\tgasfree}{t_{\rm gas-free}}

\title[Globular clusters and high-$z$ luminosity functions]
{
The Little Engines That Could? Globular Clusters Contribute Significantly to
Reionization-era Star Formation
}
\author[M. Boylan-Kolchin]
{Michael Boylan-Kolchin\\
$\!$Department of Astronomy, The University of Texas at Austin,
2515 Speedway, Stop C1400, Austin, TX 78712-1205, USA; 
\href{mailto:mbk@astro.as.utexas.edu}{mbk@astro.as.utexas.edu}}

\date{Draft version, \today}

\pubyear{2017}

\begin{document}
\label{firstpage}
\pagerange{\pageref{firstpage}--\pageref{lastpage}}
\maketitle

\defcitealias{boylan-kolchin2017}{B17}

\begin{abstract}
  Metal-poor globular clusters (GCs) are both numerous and ancient, which
  indicates that they may be important contributors to ionizing radiation in the
  reionization era. Starting from the observed number density and stellar mass
  function of old GCs at $z=0$, I compute the contribution of GCs to ultraviolet
  luminosity functions (UVLFs) in the high-redshift Universe ($10 \ga z \ga
  4$). Even under absolutely minimal assumptions -- no disruption of GCs and no
  reduction in GC stellar mass from early times to the present -- GC star
  formation contributes non-negligibly to the UVLF at luminosities that are
  accessible to the \textit{Hubble Space Telescope} (\hst;
  $M_{1500} \approx -17$). If the stellar masses of GCs were significantly
  higher in the past, as is predicted by most models explaining GC chemical
  anomalies, then GCs dominate the UV emission from many galaxies in existing
  deep-field observations. On the other hand, it is difficult to reconcile
  observed UVLFs with models requiring stellar masses at birth that exceed
  present-day stellar masses by more than a factor of 5. The \textit{James Webb
    Space Telescope} (\jwst) will be able to directly detect individual GCs at
  $z \sim 6$ in essentially all bright galaxies, and many galaxies below the
  knee of the UVLF, for most of the scenarios considered here. The properties of
  a subset of high-redshift sources with $-19 \la M_{1500} \la -14$ in \hst\
  lensing fields indicate that they may actually be GCs in formation.
\end{abstract}


\section{Introduction} 
\label{sec:intro}
The physics of galaxy formation at high redshift persists as one of the most
important questions in astrophysics. While many models exist of the ``first
galaxies'' at $z\sim20$ and putative sources of cosmic reionization at $z\sim8$
(e.g., \citealt{bromm2011, madau2017}), observational constraints are much more
challenging, especially for the lowest luminosity sources. This is a main
motivation for \jwst, yet even \jwst\ will not reveal the faintest star-forming
galaxies in the high-redshift Universe: the \textit{stellar fossil record} of
dwarf galaxies in the Local Group demonstrates the existence of galaxies in the
reionization era that are at least 40,000 times fainter than \jwst's detection
threshold in blank fields \citep{boylan-kolchin2015, weisz2017}. The integrated
ultraviolet (UV) photon output from faint but numerous galaxies is likely
crucial for maintaining an ionized intergalactic medium at $z \sim 6-8$, as
bright galaxies simply do not produce enough UV emission to maintain
reionization based on our current understanding of their escape fractions
($\fesc$; \citealt{kuhlen2012a, robertson2013, bouwens2016, finkelstein2016,
  stark2016}).

While the use of the stellar fossil record for contextualizing dwarf galaxies in
the reionization era is a relatively recent phenomenon, the basic idea of using
the fossil record to learn about the early Galaxy and its evolution has a long
and storied history \citep{eggen1962, peebles1968, searle1978, larson1990,
  bland-hawthorn2000}. Globular clusters, in particular, have captured the
attention of observers and theorists alike as they try to unravel the mysteries
of the early stages of galaxy formation. The Milky Way is surrounded by over 100
old GCs \citep{harris1996}, and their high stellar densities, characteristic
luminosity, and old ages have been the source of intrigue and speculation for
decades (for reviews, see \citealt{harris1991, vandenberg1996, brodie2006,
  charbonnel2016a}).

The typical Milky Way GC has a present-day stellar mass of $2\times 10^5\,\msun$
and an age of $12-13$ Gyr, placing its formation epoch at or near the
reionization era. The sheer number density of GCs ($n_{\rm GCs}$), coupled with
their roughly coeval formation times, points to their potential importance for
cosmic reionization: using $n_{\rm GCs}(z\!=\!0)\!=\!2\,\mpc^{-3}$
(\citealt[hereafter B17]{boylan-kolchin2017}; see also
\citealt{portegies-zwart2000}) and assuming all GCs formed at high redshifts
over a period of $\Delta\,t$, the star formation rate from GC stars is
$\dot{\rho}_{\rm \star, GC}=5\times 10^{-4}\,\xi\,\left({\Delta t/1000\,{\rm
      Myr}}\right)^{-1} \msun\, {\rm yr^{-1} \,Mpc^{-3}}\,,$ where $\xi$ is the
ratio of $\mstar$ at birth to present day. This star formation rate is a
non-negligible fraction of what is needed to maintain an ionized intergalactic
medium at $z=6$,
$\dot{\rho}_{\rm \star, crit} \approx 0.012 \,\msun\, {\rm yr^{-1} \,Mpc^{-3}}$
(\citealt{shull2012}, assuming a clumping factor of 3 and an escape fraction of
0.2). Specifically, if the mass in stars formed in GCs is a factor of $\ga 10$
larger than their present-day stellar mass, GCs can contribute $\ga 50\%$ of the
requisite ionizing flux \citep{ricotti2002, schaerer2011,
  katz2014}. \citet{carlberg2002} made the related point that the sky density of
GCs is likely to be enormous: he estimated $10^7\,{\rm deg}^{-2}$ at 1 nJy
($m_{\rm ab}=31.4$, corresponding to $\muv \approx -15.6$ at $z=7$).

In this paper, I consider the luminosity function (LF) of GCs in the
reionization era from a theoretical perspective, confront these models with
observed UVLFs, and comment on the resulting implications for GC formation
models, the contribution of GCs to reionization, and the observability of
high-redshift GCs with \hst\ and \jwst.

\section{Globular Clusters at High Redshift}
\label{sec:model}

In the low-redshift Universe, GC systems around galaxies have a nearly-universal
stellar mass function that is log-normal, with characteristic stellar mass of
$\sim 2\times 10^5\,\msun$ and $\log_{10}$ dispersion of $\sim 0.5-0.6$
\citep{harris1991, villegas2010}. Taking the known value of $n_{\rm GC}$ in the
local Universe, it is possible to compute the GC UVLF at high redshifts by
considering only the present-day stellar mass contained in GCs and a specified
formation epoch through the use of stellar population synthesis models. One can
also vary these assumptions, as several physical effects should affect GC
populations through cosmic time:
\begin{itemize}
\item \textbf{GC $\mstar$ at birth relative to present-day $\mstar$ ($\xi$)}:
  $\xi=1$ is an absolute minimum assumption, and most models that explain the
  multiple stellar populations observed in GCs (see, e.g., \citealt{gratton2012}
  and \citealt{renzini2015} for recent reviews) require much larger values,
  $\xi \sim 10-100$ (e.g., \citealt{decressin2007, dercole2008, conroy2011,
    denissenkov2014}; see \citealt{bastian2015} for a critical review of various
  mass-loss scenarios). Irrespective of the origin of multiple populations,
  evaporative losses from GCs due to collisional relaxation and mass loss from
  stellar evolution reduce the stellar mass of a GC over time
  \citep{spitzer1987, fall2001}. Furthermore, GCs are subject to impulsive tidal
  shocks as they orbit through the center of a galaxy if the time for the GC to
  traverse the disk or spheroid is short compared to the orbital period of GC
  stars. Such shocks irreversibly heat GC stars and can cause additional mass
  loss \citep{ostriker1972, spitzer1987, murali1997, gnedin1999}. The generic
  expectation from all models of GC evolution, therefore, is that $\xi>1$ for
  old GCs.
\item \textbf{Fraction of GCs surviving to $z=0$ ($\fsurv$)}: The processes
  mentioned above are enough to completely disrupt some GCs. In particular,
  two-body relaxation will tend to disrupt all but the most extended low-mass
  ($\mstar \la 10^5\,\msun$) GCs over a Hubble time, and tidal shocking should
  destroy some of the clusters that orbit closest to a galaxy's
  center (\citealt{chandrasekhar1942, spitzer1958, fall1977, binney1987}; see
  \citealt{prieto2008} for estimates of the relative contributions of
  collisional, stellar evolution, and disruption).
\item \textbf{Duration of GC formation epoch ($\Delta t$)}: While the absolute
  ages of blue GCs are relatively uncertain, it is clear that (1) blue GCs are
  ancient, with formation times comparable to the age of the Universe, and (2)
  the relative age uncertainties are significantly smaller, with a dispersion of
  approximately 0.4-0.5 Gyr (e.g., \citealt{marin-franch2009,
    vandenberg2013}). I therefore adopt $\Delta t=1000\,{\rm Myr}$ as my default
  assumption; I also consider the implications of other values of $\Delta
  t$. 1000 Myr is approximately the time between redshifts 10 and 4 (lookback
  times between 13.3 and 12.3 Gyr) in the standard cosmological model
  \citep{planck2015}. I will assume that the formation rate of globular clusters
  is a power law function of time, $\dot{n}_{\rm GCs} \propto t^{-(1+\eta)}$,
  with $\eta=-1$ (a formation rate that is constant with time) as my fiducial
  model. Note that under this assumption,
  $\dot{n}_{\rm GCs}={n}_{\rm GCs}/\Delta t$.
\end{itemize}

Considering only stars still present in $z=0$ GCs will result in a lower limit on
the contribution of GCs to high-$z$ UVLFs. The effect of $\xi$ is a uniform
shift of the entire initial GC stellar mass function to higher $\mstar$ values
(and therefore, to larger values of $L_{\rm UV}(t):\,L \rightarrow \xi L$),
while $\fsurv$ and $\Delta t$ serve to change the overall normalization of the
GC UVLF at early times:
$\phi \rightarrow \phi \, (\fsurv\,\Delta t / 1 \,{\rm Gyr})^{-1}$.

Additionally, it is not clear whether the observed log-normal LF of old GCs at
$z=0$ is reflective of their stellar mass distribution at birth versus a result
of dynamical processes operating to preferentially destroy low-mass clusters. In
the former case, the log-normal shape could be a manifestation of the Jeans mass
in the low-metallicity, high-$z$ Universe \citep{peebles1968, fall1985} or the
result of preferential disruption of low-mass proto-GC gas clouds via supernova
feedback \citep{parmentier2007, baumgardt2008}. In the latter case, the mass
spectrum of GCs at formation in the high-$z$ Universe would resemble the
spectrum of both giant molecular clouds \citep{solomon1987, rosolowsky2005,
  krumholz2015} and young star clusters forming in low-redshift systems
(especially in ongoing or recent galaxy mergers; \citealt{whitmore1995,
  larsen2002, portegies-zwart2010}); the subsequent evolution to the observed
log-normal shape is expected based on the sub-Hubble-time two-body-relaxation
timescales for low-mass clusters \citep{fall2001, prieto2008}. I will take the
initial GC stellar mass function to be log-normal as my default assumption, as
using a Schechter-like luminosity function will only result in more GCs and more
UV emission at faint magnitudes. However, I explicitly consider the case of a
Schechter-like initial stellar mass function and compare to the log-normal case
in Appendix~\ref{sec:append_schech}.

\subsection{Intrinsic Luminosity Functions}
\label{subsec:phi_intrins}
\begin{figure*}
 \centering
 \includegraphics[width=\columnwidth]{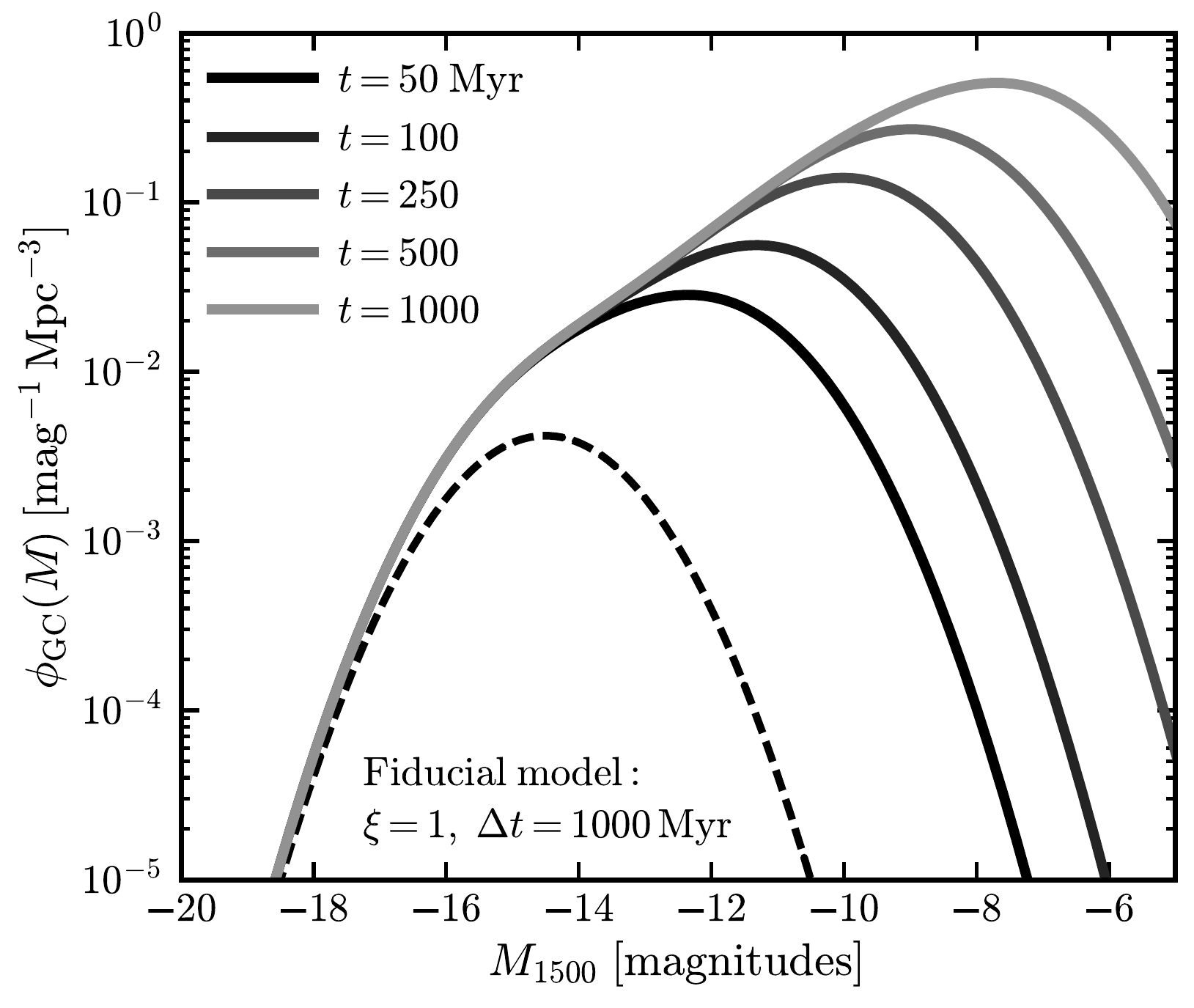}
 \includegraphics[width=\columnwidth]{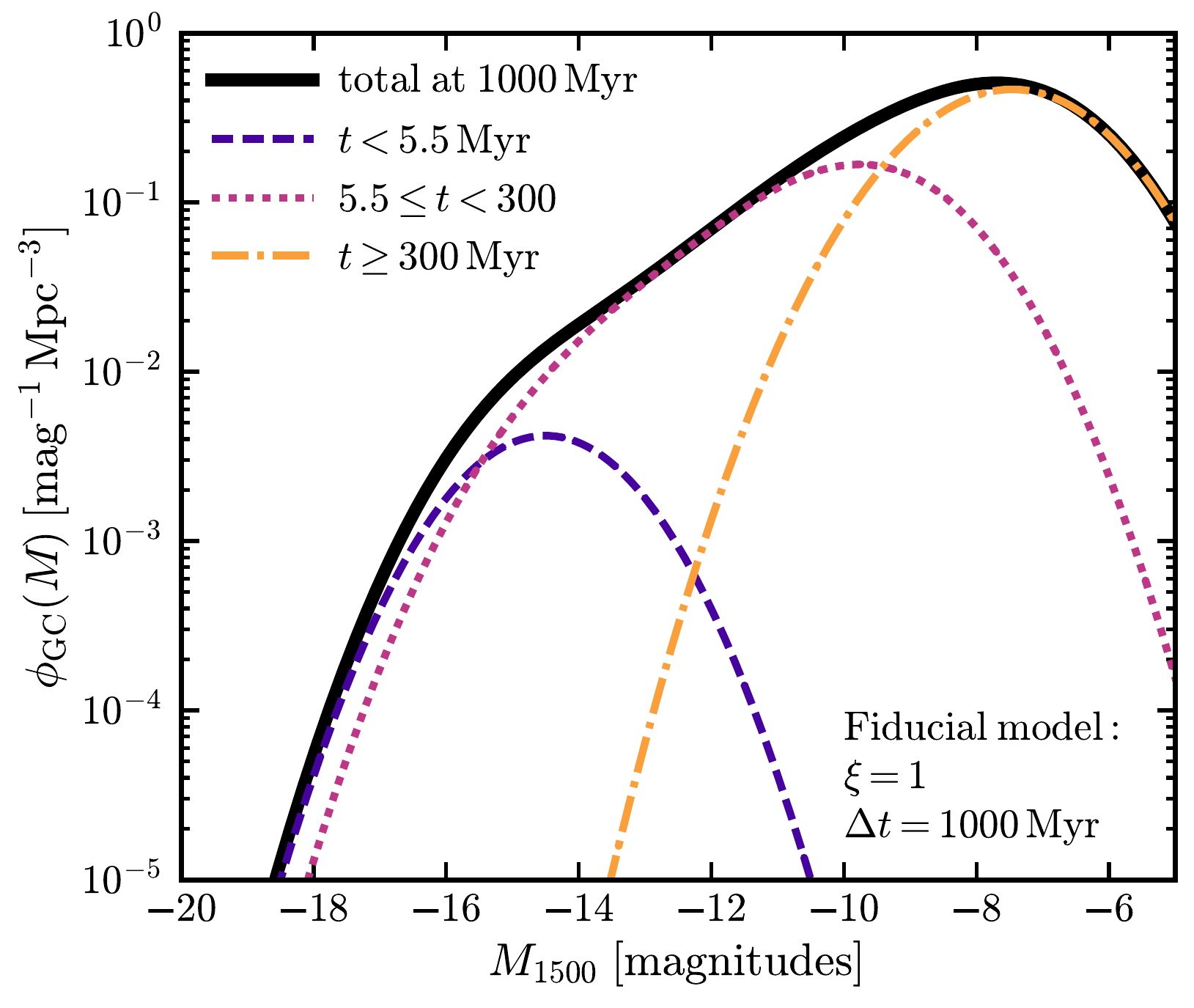}
 \caption{Intrinsic luminosity functions of GCs (uncorrected for any internal
   extinction). GCs are assumed to have $\xi=1$ and to form at a constant rate
   in time with a formation epoch duration of 
   1000 Myr. \textit{Left}: GC UVLFs at 5.5, 50, 100, 250, 500, and 1000 Myr
   after the start of the GC formation era. While GCs sample a log-normal
   luminosity function, the combination of passive fading plus continuous GC
   formation results in a total GC UVLF that develops a power-law tail with a
   cut-off at faint magnitudes. \textit{Right}: The contribution of GCs with
   various ages to the GC UVLF at 1000 Myr. While the youngest GCs (with ages
   $<5.5$ Myr) are the brightest, the GC UVLF is dominated by GCs formed in the
   previous 5.5--300 Myr.
 \label{fig:lf_onlyGCs}
}
\end{figure*}

Assuming a log-normal form for the $z=0$ GC stellar mass function, the UVLF of a
cosmologically representative GC system (measured at 1500~\AA\ in the rest
frame, $M_{1500}$) can be thought of a log-normal distribution with a mean value
$\overline{M}$ that evolves with time. $\overline{M}(t)$ is fully determined by
the time evolution of UV emission from a specified stellar population and the time
dependence of any mass loss; based on stellar population synthesis calculations
outlined in \citetalias{boylan-kolchin2017}, I adopt
\begin{flalign}
\overline{M}(t)&=-14.5-2.5\,\log_{10}\left[\xi\,a\,\left(\frac{t}{\rm
      Myr}\right)^{-b}\right],\;{\rm with}
\label{eq:m1500}\\
(a, \,b)&=\begin{cases}\label{eq:t_var}
(1, 0)\; &{\rm if } \;t<5.5\,{\rm Myr}\\
(9.17, 1.3)\,& {\rm if} \;5.5\le t/{\rm Myr} \la 300\\
(103, 1.725)\,& {\rm if} \;300\le t/{\rm Myr} \la 1400
\end{cases}
\end{flalign}

Fig.~\ref{fig:lf_onlyGCs} shows the evolution of the GC UVLF over the full epoch
of GC formation (see Appendix~\ref{sec:append_lognorm} for details of this
calculation). Initially, the GC UVLF traces the log-normal form of the
underlying GC mass function. As time progresses, however, two competing effects
take place: the earliest-forming GCs fade passively, shifting uniformly to
fainter magnitudes, and newly-forming GCs emerge, populating the initial
log-normal GC UVLF curve. The full GC UVLF is a sum over populations of GCs of
different ages (and, correspondingly, different mean $\muv$ values).  The shape
of the full GC UVLF is therefore determined by how quickly $\muv$ fades with
time for a given GC: for sufficiently fast fading, only the very early stages of
GC formation will be important for the UVLF, while GCs of all ages contribute
roughly equally in the limit of very slow fading.

The relative contributions to the GC UVLF at the end of the GC formation epoch
(which is assumed to last 1 Gyr), for the periods $t < 5.5$ Myr, $5.5 < t < 300$
Myr, and $t > 300$ Myr, are shown in the right panel of
Figure~\ref{fig:lf_onlyGCs}. While the very brightest portion of the GC UVLF is
dominated by the youngest GCs, it is GCs with $5.5 < t < 300$ Myr that
contribute the bulk of the UVLF for a wide range of luminosities, as the number
of GCs in this age range is a factor of 50 larger than the newly-formed GCs. The
oldest GCs contribute only to the faintest portion of the LF. This is directly
related to the choice of a \citet{kroupa2001} stellar initial mass function
(IMF), for which the luminosity of a GC evolves as $L \propto t^{-b}$, with the
time-dependence of $b$ noted in Eq.~\eqref{eq:t_var}.

\begin{figure}
 \centering
 \includegraphics[width=\columnwidth]{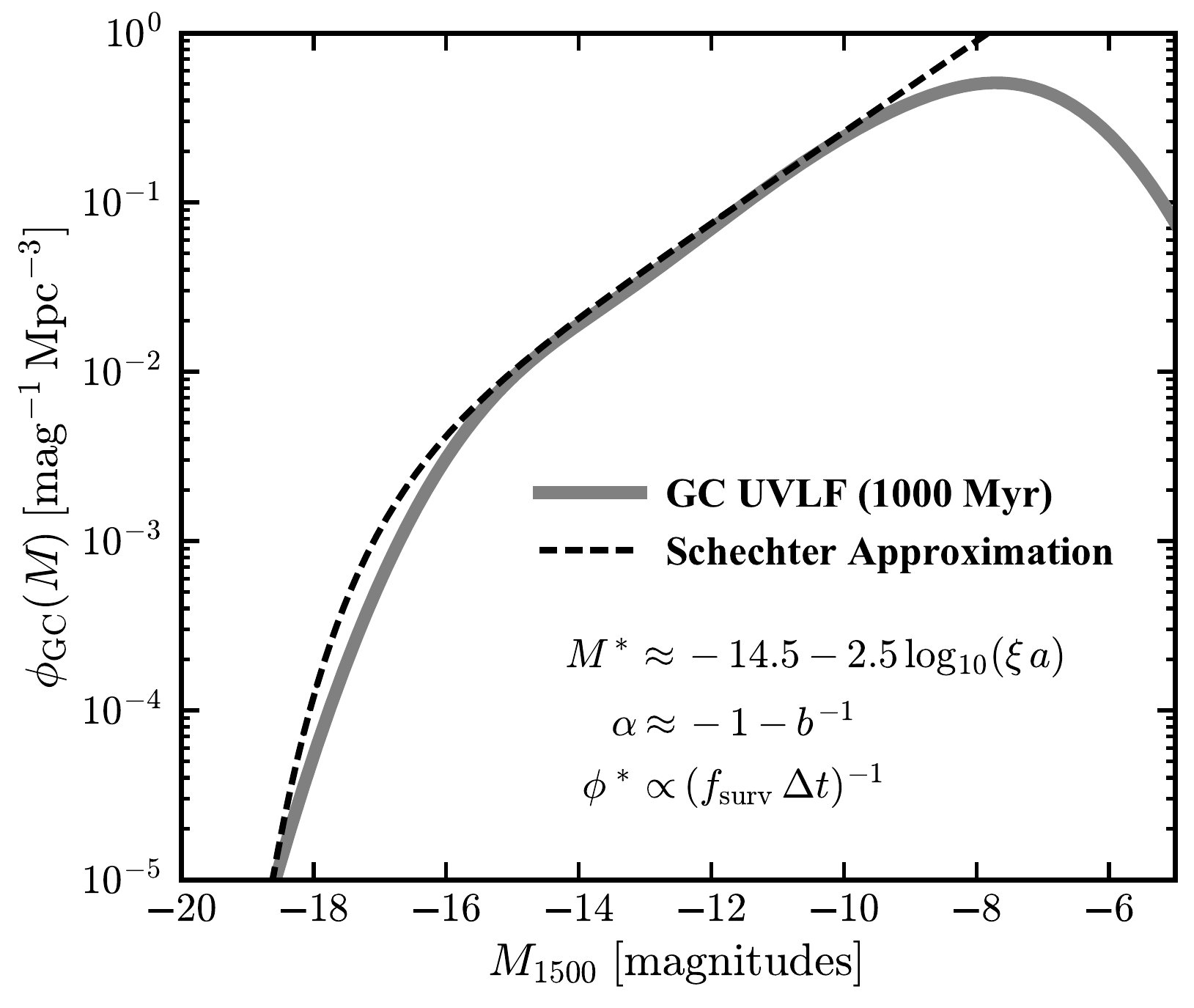}
 \caption{Luminosity function from GCs in the fiducial model (in which the GC
   stellar mass function at birth is log-normal) at $t=1000$ Myr (solid gray),
   along with Schechter function approximation (dashed black).  The parameters
   of the Schechter function ($M^*, \,\alpha, \,\phi^*$) are determined by
   $\xi$, $\Delta t$, and stellar evolution (via $a,\,b$; see
   Eq.~\eqref{eq:m1500}).
 \label{fig:schechter}
}
\end{figure}

Intriguingly, even though the underlying UVLF of GCs forming at any particular
instant is log-normal, the \textit{overall} GC UVLF can be approximated by a
\citet{schechter1976} function for $t \gg 5.5$ Myr
(Figure~\ref{fig:schechter}). The faint-end slope $\alpha$ of this Schechter-like
LF is determined only by the fading of a GC's UV luminosity with time,
$\alpha \approx -1-1/b$, while the characteristic magnitude is given by
$M^*=-14.5-2.5\,\log_{10}(\xi\,a)$. Since the GC UVLF is dominated by GCs with
ages of $5.5 < t < 300$ Myr, it is the values of $a$ and $b$ corresponding
to this period that are appropriate: $a=9.17$ and $b=1.3$, yielding
$\alpha \approx -1.77$ and $M^* \approx -16.9-2.5\,\log_{10} \xi$. In
Appendix~\ref{sec:append_schech}, I present analogous results for GC stellar
mass functions that are themselves Schechter-like at formation.

\subsection{Contributions of GCs to $\phi_{\rm UV}$}
\begin{figure*}
 \centering
\includegraphics[width=0.7\textwidth]{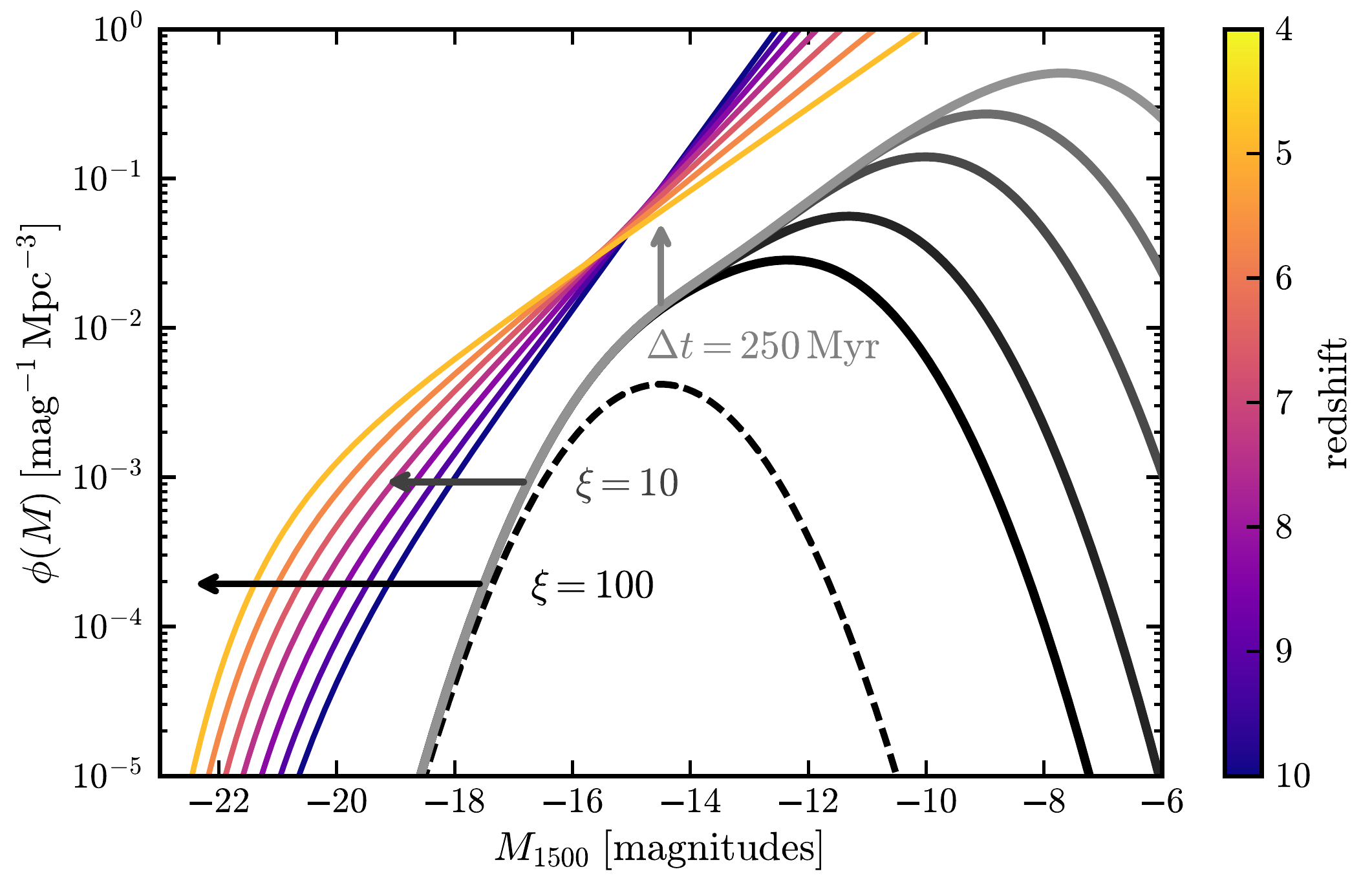}
\caption{The intrinsic luminosity function of globular clusters (gray-scale,
  with line colors identical to those in the left panel of
  Fig.~\ref{fig:lf_onlyGCs}) during their epoch of formation. The plot assumes
  birth masses are equal to present-day masses ($\xi=1$) and formation
  uniformly distributed in time over a period of $\Delta t=1$ Gyr, approximately
  the time from $z=10$ to $z=4$. Also plotted are
  observed UV LFs (colored lines) from $z=10$ (dark purple) to $z=4$ (light orange) as
  compiled in \citet{finkelstein2016}. Horizontal arrows indicate the effects of
  assuming $\xi=10$ (dark gray) or 100 (black), while the vertical arrow shows the
  shift in the globular cluster luminosity function if the formation period is
  250 Myr. This would also be the shift required if $\fsurv=0.25$ as opposed to
  1. 
 \label{fig:lf_vs_fink}
}
\end{figure*}
Figure~\ref{fig:lf_vs_fink} compares the intrinsic (unobscured) GC UVLF to
measured global UVLFs from $z=10$ to $z=4$ (as compiled in \S 5.2-5.3 of
\citealt{finkelstein2016}). For magnitudes fainter than $\muv \approx -17$,
there is significant uncertainty in the faint-end slope of the UVLF,
particularly at $z \ga 8$. The plotted results assume steeper faint-end slopes
compared to some estimates and therefore can be interpreted as providing upper
limits to the luminosity function of faint galaxies at early times. For the
default assumptions adopted here --
$\xi=1,\;\Delta t=1\,{\rm Gyr},\;{\rm and }\;\fsurv=1$ -- GCs contribute a
maximum of $\approx 25\%$ of the UVLF (at $\muv \approx -15$), with a lesser
contribution for very bright magnitudes (because GCs have a maximum initial
stellar mass) and very faint magnitudes (because the observed high-$z$ UVLFs
have very steep faint-end slopes) at most redshifts.

The inferred contribution of GCs to the global UVLF changes dramatically if we
consider alternate scenarios, however: the birth masses of GCs ($\xi$), GC
disruption ($\fsurv$), and the duration of the GC formation epoch ($\Delta t$)
all have significant effects on the contribution of GCs to the global UVLF. The
horizontal location of the GC UVLF depends on $\xi$ while its normalization is
proportional to $(\fsurv\,\Delta t)^{-1}$. These effects are noted with arrows
in Fig.~\ref{fig:lf_vs_fink} and indicate that $\xi=100$ is extremely difficult
to reconcile with observed luminosity functions.

Having established the unobscured GC UVLF, the obvious next question is: do GCs
contribute significantly to the global UVLF at high redshifts? To address this
question, I need to convert from intrinsic to observed UV fluxes, i.e., to
account only for UV ({rest-frame 1500 \AA) radiation that escapes the GC. A
  related quantity -- the fraction of \textit{ionizing} radiation that escapes
  the galaxy, $\fesc$ -- is perhaps the most fraught aspect of reionization
  modeling. Many global models of reionization assume a constant escape fraction
  of $\fesc \approx 0.1-0.2$ (e.g., \citealt{robertson2013, finkelstein2015,
    ishigaki2017}). No such consensus exists in simulations, with results
  ranging from fairly high to virtually negligible values of $\fesc$ (e.g.,
  \citealt{wise2014, paardekooper2015, ma2015, xu2016, gnedin2016, anderson2017,
    howard2017a, trebitsch2017, kimm2017, zackrisson2017}). However, the escape
  fraction of UV radiation (hereafter, $\fescobs$) will certainly be higher than
  $\fesc$, the escape fraction of hydrogen-ionizing radiation.

\begin{figure*}
  \centering
  \includegraphics[width=0.7\textwidth]{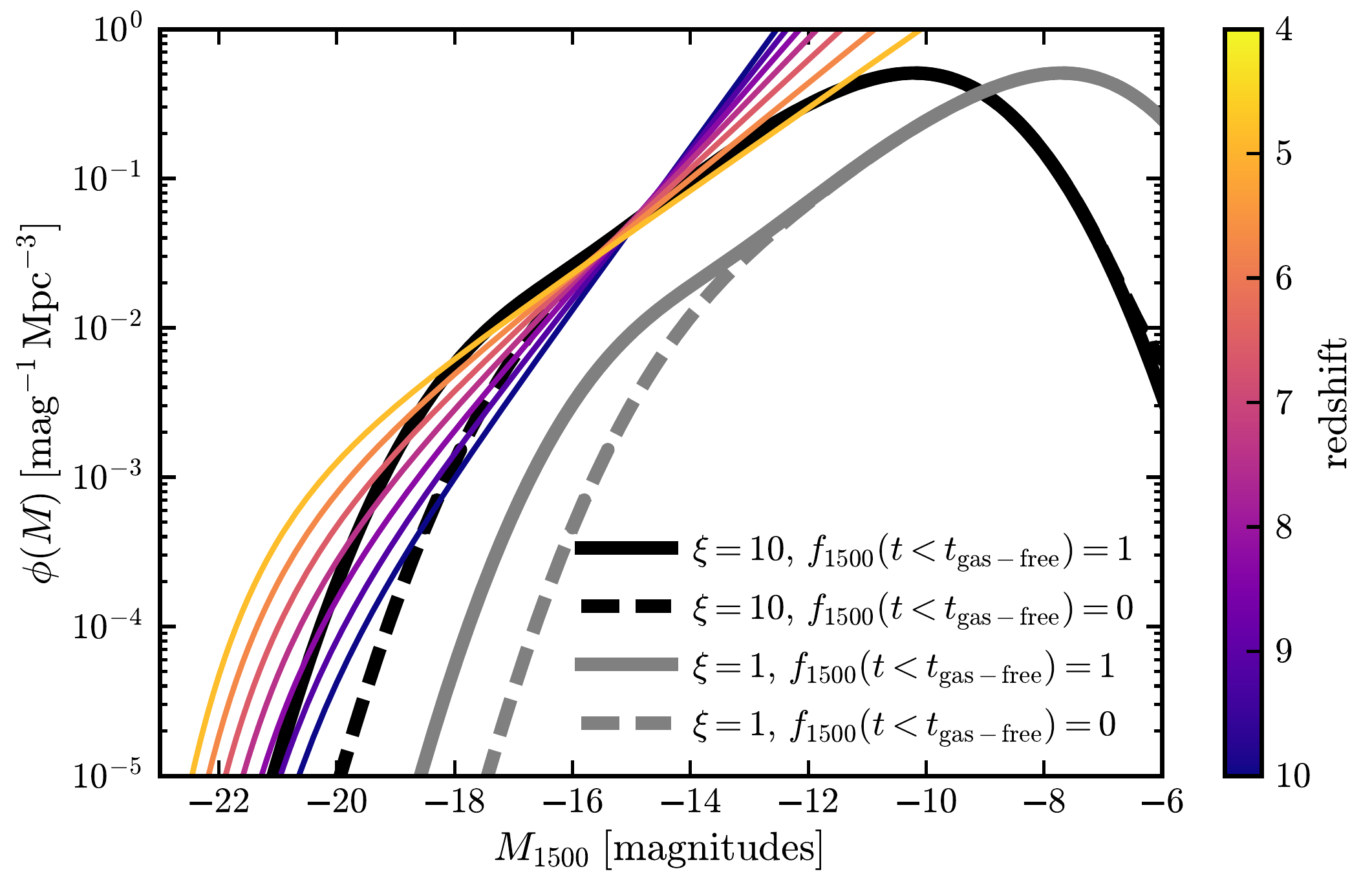}
  \caption{Observed UV LFs (colored lines) from $z=10$ to $z=4$ (as in
    Fig.~\ref{fig:lf_vs_fink}, along with modeled observed GC UVLFs (thick black
    and gray lines). The black lines assume $\xi=10$, while the gray lines
    assume $\xi=1$; solid lines assume $\fescobs (t<\tgasfree)=1$, while dashed
    assume $\fescobs (t<\tgasfree)=0$ (here, $\tgasfree=10$ Myr). All cases
    assume $\fescobs (t>\tgasfree)=1$; different values of $\fescobs$ can be
    folded into $\xieff \equiv \xi\,\fescobs$ (see text for details). Models
    with $\xi=10$ exceed observed UVLFs at levels that would likely have been
    already detected. Assuming that no UV radiation can escape from GCs at early
    times $(t<\tgasfree \approx 10\,{\rm Myr})$ affects the bright end of
    $\phi(M)$ but leaves it unaffected at fainter magnitudes.
 \label{fig:lf_obs_vs_fink}
}
\end{figure*}
Observations of young massive clusters reveal that they are gas-free within
$\sim 5-10$ Myr of their birth \citep{bastian2014, hollyhead2015}, indicating
that a combination of stellar winds, radiation pressure, and type II supernovae
is highly efficient at completely removing gas from nascent GCs. It is therefore
likely that $\fesc$ for GCs depends time relative to $\tgasfree$, where
$\tgasfree=10$ Myr is the assumed timescale for the evacuation gas from a
newly-formed GC; this value is a conservative upper
limit. Furthermore, the distribution of GCs within their host halos is
significantly more extended than the size of the central galaxy
\citep{hudson2017, forbes2017}, indicating that GCs typically form in regions
that, while locally very dense, are globally embedded in a relatively tenuous
medium from which photons can easily escape.

I will therefore consider two possibilities: $\fescobs=1$ at all times, or
$\fescobs (t<\tgasfree)=0$ and $\fescobs (t>\tgasfree)=1$ (corresponding to no
1500 \AA\ UV escape prior to gas evacuation and full escape afterward). Lower
values of $\fescobs$ can be modeled by shifting
$M_{1500} \rightarrow M_{1500}-2.5\log_{10}(\fescobs)$. I consider both $\xi=1$
and 10 for each $\fescobs$ scenario. While the $\fescobs=0$ scenario is
unrealistic, it provides a firm lower limit and therefore serves a useful role
in demonstrating the lowest possible contribution of GCs to UVLFs.

Figure~\ref{fig:lf_obs_vs_fink} compares the GC UVLFs computed using these
parameters to the same galaxy UVLFs from Fig.~\ref{fig:lf_vs_fink}. Black curves
assume $\xi=10$ and gray curves assume $\xi=1$; solid and dashed lines
correspond to $\fescobs (t<\tgasfree)=1$ and 0, respectively. The models with
$\xi=1$ fall below observations, while both models with $\xi=10$ exceed observed
UVLFs at magnitudes accessible to \hst\ in blank fields ($M_{1500} \la -17$;
\citealt{mclure2013, schenker2013}). However, I continue to consider these models with
$\xi=10$ for a number of reasons. First, $\fescobs$ and $\xi$ can
be combined into a single parameter, $\xieff \equiv \xi \, \fescobs$; this is
the quantity that is constrained by observations and the gray (black) lines in
Fig~\ref{fig:lf_obs_vs_fink} are best thought of as lines with $\xieff =1$
(10). Second, $\xi \ga 10$ is routinely invoked as a necessity in literature
models of light element anti-correlations in GCs. Finally, high values of $\xi$
provide a firm upper limit to the number of GCs that should be detectable in the
high-redshift Universe. In both cases ($\xi=1$ and 10), the faint-end slope of
the GC UVLF is very similar to the $z=4$ UVLF ($\alpha \approx -1.7$). A clear
prediction, independent of $\xi$, is that the GC UVLF should extend to high
number densities at faint magnitudes: the LF turn-over occurs at
$M_{1500} \approx -7.75$ ($-10.25$) at the end of the GC formation epoch for
$\xi=1$ (10). The time evolution of the turn-over magnitude can be seen in
Figs.~\ref{fig:lf_onlyGCs}~and~\ref{fig:lf_vs_fink}.

Figure~\ref{fig:ratio_obs_vs_fink} compares the GC UVLFs to global UVLFs in
ratio form in order to assess the contribution of GCs to UVLFs
explicitly. Dashed lines correspond to $\xieff=10$ and solid lines correspond to
$\xieff=1$; in each case, I assume $\fescobs (t<\tgasfree)=0$ and 1 at later
times. For large values of $\xieff$, a large fraction of the global UVLF comes
from GCs (in fact, the GC UVLF exceeds the total UVLF at observable magnitudes
for $\xieff=10$). The contribution of GCs is substantially lower for $\xieff=1$,
particularly for magnitudes brighter than \hst's blank-field limit of
$M_{1500} \approx -17$. At fainter magnitudes ($M_{1500} \ga -15$), the
contribution of GCs to the UVLF approaches or exceeds 10\% at all redshifts. If
the faint-end slope of the galaxy UVLF is shallower than the estimates of
\citet{finkelstein2016}, the relative contribution of globular clusters would be
enhanced relative to what is plotted.

It therefore appears unavoidable that GCs contribute appreciably (at minimum) or
perhaps dominantly (for high values of $\xieff$) to high-redshift
UVLFs. \citet{katz2013} reached similar conclusions with a complementary
approach to computing high-$z$ contributions of GCs to UVLFs. Their derived GC
UVLFs are significantly steeper than those obtained here, however: even though
their assumed low-mass slope of the initial GC stellar mass function is $-1.5$,
\citet{katz2013} find GC UVLFs that have faint-end slopes that are steeper than
$-2$. As is shown in Appendix~\ref{sec:append_schech}, the GC UVLFs calculated
here have slopes that are shallower than $-2$ even if the initial GC stellar
mass function is assumed to have a slope of $-1.9$.

\section{Discussion}
\label{sec:highz}
The results of the previous section indicate that GCs can contribute
substantially to high-redshift UV emission. A natural question, therefore, is
whether individual GCs will be detectable by future (or even current) facilities, or
if only the sum of the light from multiple GCs will be visible. 

Figure~\ref{fig:montecarlo} shows the probability distribution of $M_{1500}$
from the integrated light in GCs for systems with 1, 10, 100, and 1000 GCs at
$z=6$. The plotted distributions assume $\xieff=1$ and
$\fescobs (t<\tgasfree)=0$; results for other values of $\xieff$ can be obtained
through a straightforward rescaling (as noted on the $x$-axis of the
figure). Changing from $\fescobs (t<\tgasfree)=0$ to 1 would result in a
brightward shift of 0.6 magnitudes for each distribution, as $\approx 40\%$ of
the lifetime UV emission of each GC comes in the first 10 Myr based on the
assumptions adopted here. If $\tgasfree$ is set to 5 Myr rather than 10 Myr,
then 25\% of the lifetime UV emission is obscured, corresponding to a brightward
shift of 0.3 mag.

Whether or not any individual GC is detectable depends on how likely it is to
exceed the detection threshold of a given instrument. Table~\ref{tab:table1}
lists the probability that at least one GC can be directly detected in halos of
various masses at $z \approx 6$ with \hst\ or \jwst. The $\mhalo-M_{1500}$
relationship is computed via abundance matching the halo mass function (computed
following \citealt{sheth2001} using the \citealt{planck2015} cosmology) and the
observed $z\sim 6$ UVLF. Results are listed separately for models with
$\xieff=1$ (models 1a and 1b) and $\xieff=10$ (models 2a and 2b). For the
rarest, most massive halos at $z=6$ ($\mhalo \approx 10^{12}\,\msun$,
corresponding to $M_{1500} \approx -21.6$), \hst\ will be able to detect at
least 1 GC 2-20\% of the time if $\xieff=1$ and in virtually every case if
$\xieff=10$. \jwst\ can detect at least one GC directly in essentially all
scenarios considered here. Moving to lower-mass halos, detection becomes
increasingly unlikely. Nevertheless, \jwst\ will detect one GC in 1.5\% (39\%)
of halos with $\mhalo(z\sim 6) \approx 10^{10}\,\msun$ (corresponding to
$M_{1500} \approx -15.6$, which is near the faint limit of \jwst\ detections in
hypothetical deep fields) under pessimistic (optimistic) assumptions.

The possibility of detecting GCs, or that some current observations may already
be revealing proto-GCs, in the high-redshift Universe was anticipated by
\citet{carlberg2002} and has recently gained significant attention
\citep{schaerer2011, katz2013, bouwens2017, boylan-kolchin2017, renzini2017,
  vanzella2017}. In particular, the very small sizes (10s of pc) of
intrinsically faint, gravitationally lensed sources at $z\sim 6$ in the
\textit{Hubble} Frontier Fields appear consistent with GCs or star cluster
complexes in formation \citep{kawamata2015, vanzella2017, bouwens2017,
  kawamata2018}: figure~12 of \citet{bouwens2017} shows a population of galaxies
with roughly constant sizes of $5-15$ pc over the luminosity range
$-19.5 \la M_{1500} \la -14$. GCs themselves are likely easier to detect than
other objects with similar values of $M_{1500}$, as their surface brightnesses
are substantially higher (see \citealt{zick2018} for a more detailed
discussion).

Compact, UV-bright sources are also found in simulations of reionization-era
galaxies \citep{kimm2016, ma2017a, kim2018}. The long-term stability of UV
clumps is unclear: some may be clusters that are not gravitationally bound and
disrupt rapidly after formation (within $\sim 10^7$ years), similar to star
clusters in low-redshift merging systems \citep{bastian2005,fall2005}. If this
is the case, an even higher fraction of high-redshift star formation may occur
in dense, GC-like systems than the estimates in this paper imply, as these
disrupting clusters would not evolve into GCs at $z=0$. In fact, it may be
possible to model high-$z$ UVLFs using the tools presented here in the limit
that all stars form in clusters with either log-normal or Schechter-like shapes.

There is certainly room for improvement in the modeling presented here. Perhaps
the most important refinement would be to include a $\mstar$ dependence in
$\xieff$; detailed estimates of $\fsurv$ would also be highly informative. A
better understanding of both absolute and relative ages for blue GCs is also
essential to model their relevance in the reionization era and to constrain
$\Delta t$ (though it appears difficult to reconcile a later formation epoch
with observed properties of $z \sim 4$ galaxies; \citealt{katz2013}). The
stellar IMF is also a source of uncertainty: more top-heavy IMFs will produce
more UV luminosity per unit stellar mass, changing the relationship between
$M_{1500}$ and $\mstar$. Nonetheless, the power of the relatively
straightforward models described in this paper to directly constrain properties
of GCs in the high-redshift Universe and their contributions to cosmic
reionization is encouraging. Some ``galaxies'' already detected in \hst\ deep
fields are likely to be GCs in formation; in the \jwst\ era, observations of GCs
in formation should be commonplace.

\begin{figure}
 \centering
\includegraphics[width=\columnwidth]{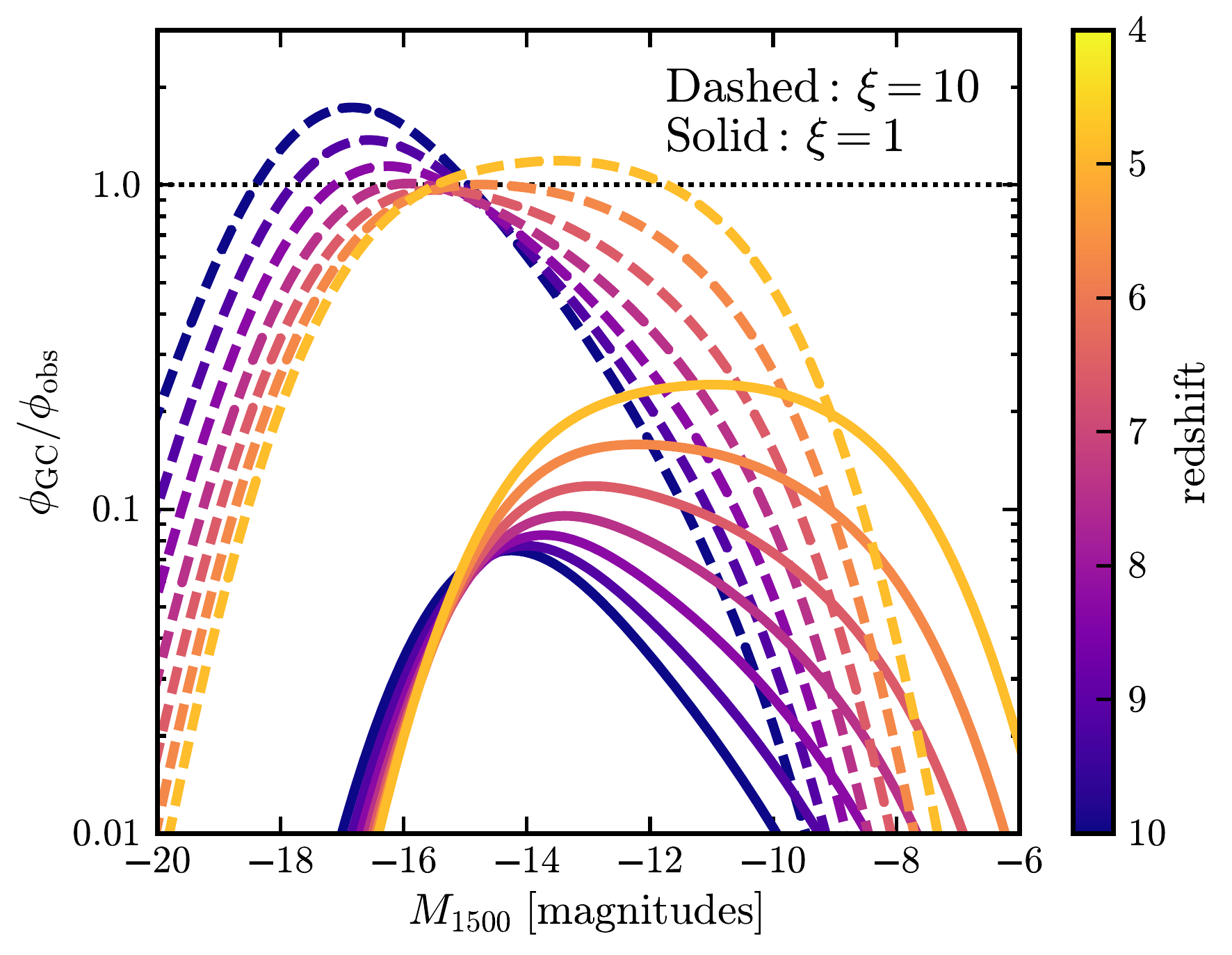}
\caption{Ratio of GC UVLFs to global UVLFs at a variety of redshifts under the
  assumption that $\xi=1$ (solid lines) or 10 (dashed lines), assuming
  $\fescobs=0$ for $t<\tgasfree$ and 1 otherwise. If $\xi=1$, then 
  $\la 1\%$ of the UVLF accessible to \hst\ in blank fields ($M_{1500} < -17$)
  originates from GCs; by $M_{1500} \approx -14$, approximately 20\% of the UVLF
  comes from GCs. The GC UVLF actually exceeds the global UVLF at most redshifts
  at magnitudes accessible to \hst\ if $\xi=10$.
 \label{fig:ratio_obs_vs_fink}
}
\end{figure}
\begin{figure}
 \centering
 \includegraphics[width=\columnwidth]{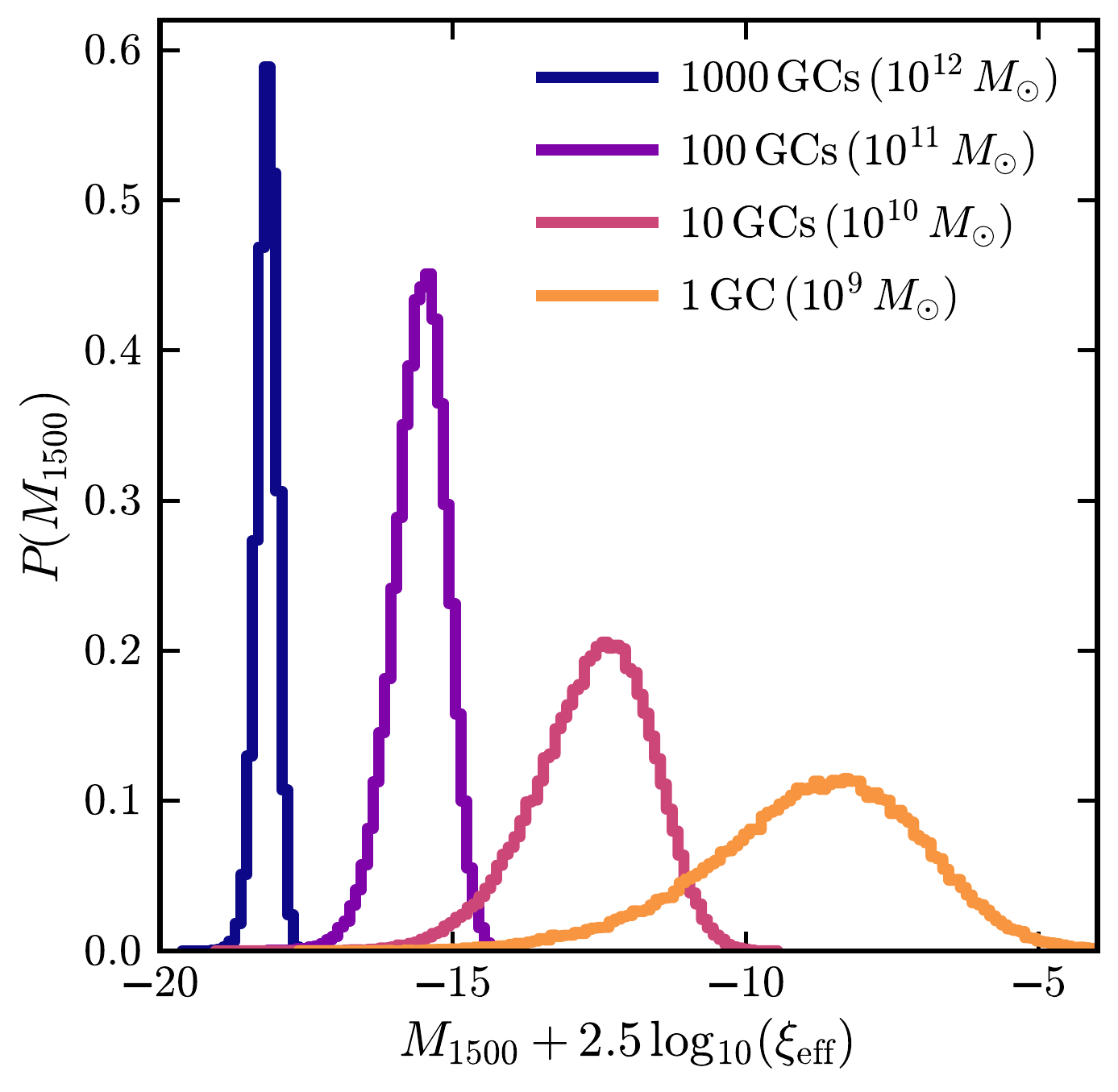}
 \caption{Probability distributions for $M_{1500}$ from GCs (integrated over a
   halo's GC population) for halos of 4 different masses at $z\sim6$. Halos at the
   threshold of GC formation ($\mhalo(z=6)=10^{9}\,\msun$,
   \citetalias{boylan-kolchin2017}) span a wide range of possible values of
   $M_{1500}$, as any individual GC has a wide range of possible masses and
   formation times. At increasingly high masses, and correspondingly higher
   numbers of GCs, the widths of the distributions shrink, as the GCs come
   closer to fully sampling the underlying luminosity function. All calculations
   for the plot assume that $\xieff=1$ (though the results can easily be scaled
   to other values of $\xieff$, as is indicated on the $x$-axis) and
   $\fescobs (t< \tgasfree)=0$; the probability distribution for a
   $10^{12}\,\msun$ halo has been rescaled downward by a factor of 2 to decrease
   the dynamic range.
 \label{fig:montecarlo}
}
\end{figure}
\begin{table}
\renewcommand{\arraystretch}{1.4}
\setlength{\tabcolsep}{6pt}
\caption{\textit{Observability of GCs at high redshifts.} Model 1 assumes
  $\xieff=1$, while model 2 assumes $\xieff=10$. Model $a$ assumes 
  $\fescobs (t<\tgasfree)=0$, while model $b$ assumes $\fescobs
  (t<\tgasfree=1)$. The second column gives $M_{1500}$ corresponding to the
  average value of the luminosity coming from all GCs in the given halo, while
  the third column gives the median value of $M_{1500}$ and the symmetric 80\%
  confidence interval about the median. Columns 4 and 5 give the fraction of
  such halos that will host at least one observable GC for \hst\ (assumed
  detection threshold of $M_{1500}=-17$) and \jwst\ (assumed detection threshold
  of $M_{1500}=-15)$. 
}
  \label{tab:table1}
  \begin{tabular}{@{   }c r r r r r@{   }} 
    \hline
    \hline
    Model & $M_{1500}$(GCs) & $M_{1500}$(GCs)& $f_{\rm obs}$ (\hst) & $f_{\rm obs}$ (\jwst) & 
    \\
\hline
    \multicolumn{6}{c}{$\boldsymbol{\mhalo}\mathbf{=10^9}\, 
    \boldsymbol{\msun}$, $\boldsymbol{M_{1500}}\mathbf{=-12.2}$}\\
1a & -10.6 & $-8.7^{+2.1}_{-2.6}$ & $1.0\times 10^{-5}$ & $1.5\times 10^{-3}$ &\\
1b & -11.2 & $-8.7^{+2.1}_{-2.7}$ & $2.3\times 10^{-4}$ & $5.5\times 10^{-3}$ &\\
2a & -13.1 & $-11.2^{+2.1}_{-2.6}$ & $3.1\times 10^{-3}$ & $0.034$ &\\
2b & -13.7 & $-11.2^{+2.1}_{-2.7}$ & $9.0\times 10^{-3}$ & $0.046$ &\\
\hline
    \multicolumn{6}{c}{$\boldsymbol{\mhalo}\mathbf{=10^{10}}\, 
    \boldsymbol{\msun}$, $\boldsymbol{M_{1500}}\mathbf{=-15.6}$}\\ 
1a & -13.1 & $-12.5^{+1.1}_{-1.4}$ & $2.0\times 10^{-4}$& $0.014$ &\\
1b & -13.7 & $-12.8^{+1.2}_{-1.9}$ & $2.6\times 10^{-3}$& $0.054$ &\\
2a & -15.6 & $-15.0^{+1.1}_{-1.4}$ & $0.032$& $0.30$ &\\
2b & -16.2 & $-15.3^{+1.2}_{-1.9}$ & $0.092$& $0.39$ &\\
\hline
    \multicolumn{6}{c}{$\boldsymbol{\mhalo}\mathbf{=10^{11}}\, 
    \boldsymbol{\msun}$, $\boldsymbol{M_{1500}}\mathbf{=-19.1}$}\\
1a & -15.6  & $-15.5^{+0.5}_{-0.6}$ & $1.9\times 10^{-3}$& $0.14$ &\\
1b & -16.2  & $-16.1^{+0.7}_{-0.9}$ & $0.026$& $0.43$ &\\
2a & -18.1  & $-18.1^{+0.5}_{-0.6}$ & $0.28$& $0.97$ &\\
2b & -18.7  & $-18.6^{+0.7}_{-0.9}$ & $0.61$& $0.99$ &\\
\hline
    \multicolumn{6}{c}{$\boldsymbol{\mhalo}\mathbf{=10^{12}}\, 
    \boldsymbol{\msun}$, $\boldsymbol{M_{1500}}\mathbf{=-21.6}$}\\
1a & -18.1 & $-18.2^{+0.2}_{-0.2}$ & $0.018$& $0.76$ &\\
1b & -18.7 & $-18.8^{+0.3}_{-0.3}$ & $0.23$& $1.0$ &\\
2a & -20.6 & $-20.7^{+0.2}_{-0.2}$ & $0.96$& $1.0$ &\\
2b & -21.2 & $-21.3^{+0.3}_{-0.3}$& $1.0$& $1.0$ &\\
    \hline
  \end{tabular}
\end{table}

\section*{Acknowledgments} 
I thank Charlie Conroy and Dan Weisz for insightful comments on earlier versions
of this paper; Michael Fall, Kristian Finlator, Oleg Gnedin, Pawan Kumar, Smadar
Naoz, Eliot Quataert, Charli Sakari, Chris Sneden, and Eros Vanzella for helpful
discussions; and Tom Petty for aural support. Support for this work was provided
by The University of Texas at Austin, the National Science Foundation (grant
AST-1517226), and NASA through grant NNX17AG29G and HST grants AR-12836,
AR-13888, AR-13896, GO-14191, and AR-14282 awarded by the Space Telescope
Science Institute, which is operated by the Association of Universities for
Research in Astronomy, Inc., under NASA contract NAS5-26555. Much of the
analysis in this paper relied on the python packages {\tt NumPy} \citep{numpy},
{\tt SciPy} \citep{scipy}, {\tt Matplotlib} \citep{matplotlib}, and {\tt
  iPython} \citep{ipython}; I am very grateful to the developers of these
tools. This research has made extensive use of NASA's Astrophysics Data System
(\href{http://adsabs.harvard.edu/}{http://adsabs.harvard.edu/}) and the arXiv
eprint service (\href{http://arxiv.org}{http://arxiv.org}).

\bibliography{draft_clean}

\appendix
\section{Log-Normal Luminosity Functions}
\label{sec:append_lognorm}
The component of the GCLF originating from GCs with ages in the range $[t_i,\,t]$
can then be written as a time integral over the individual cluster luminosity
functions:
\begin{equation}
  \label{eq:ap_phi_1}
  \phi_{\rm GC}^i(M, t) \propto
  \int_{t_i}^t \exp\left(-\frac{(M-\overline{M}(t))^2}{\sqrt{2\,\sigma^2}}\right) \, f(t)\,\,dt\,.
\end{equation}
If $f(t)\propto t^{-(1+\eta)}$, the resulting integral can be computed
analytically \citep{basu2015}:
\begin{flalign}
  \label{eq:ap_phi_2}
  \phi_{\rm GC}^i(M, t) &\propto
  \exp\left(\frac{\eta\,B}{C}+\frac{\eta^2}{4C^2}\right) \, [E_\eta(t)-E_\eta(t_i)]\,,\\
  E_\eta(\tau) &\equiv {\rm erf} 
  \left(B+\frac{\eta}{2C}+C\ln\left(\frac{\tau}{{\rm Myr}}\right)\right)
\end{flalign}
where
\begin{flalign}
  \label{eq:B}
  B&=-\frac{M-M^*}{\sqrt{2\,\sigma^2}}\,,\\
  C&=\frac{2.5\,b}{\ln(10)\,\sqrt{2\,\sigma^2}} \,,\;{\rm and}\\
  M^*&=-14.5-2.5\,\log_{10}(\xi\,a) \,.
\end{flalign}
I will assume $\eta=-1$ (i.e., $\dot{n}_{\rm GC}={\rm constant}$).  The full
intrinsic GC UVLF is then a sum over component $\phi_{\rm GC}^i(M, t)$ values,
with piecewise time ranges specified by Eq.~\eqref{eq:t_var}, appropriately
normalized for the time evolution of $\dot{n}_{\rm GC}$.

\section{Schechter Luminosity Functions}
\label{sec:append_schech}
\begin{figure}
 \centering
 \includegraphics[width=\columnwidth]{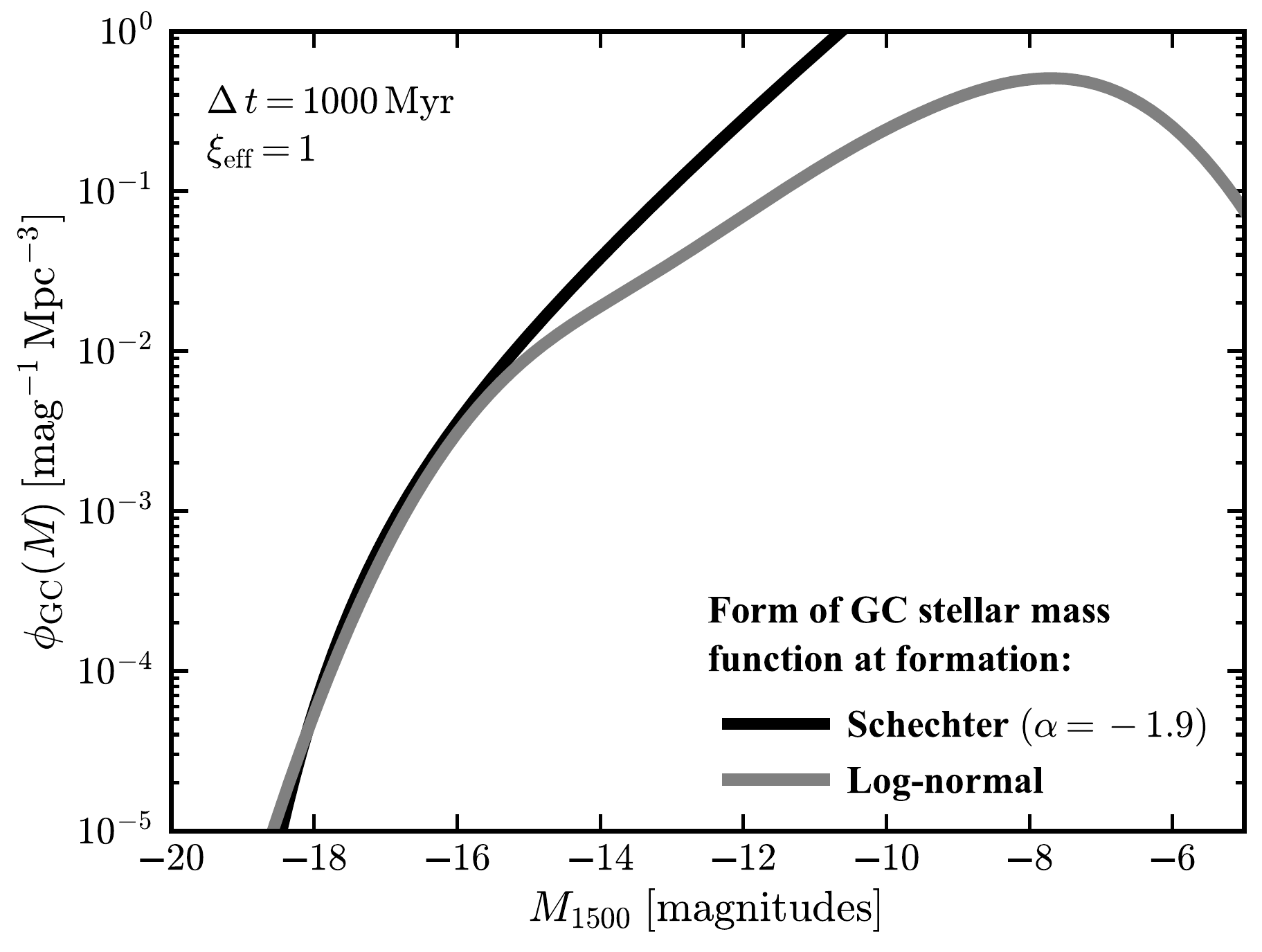}
 \caption{A comparison between the derived GC UVLF at the end of the GC
   formation epoch (after 1 Gyr) under the assumption of initially log-normal or
   Schechter-like GC stellar mass functions (gray and black, respectively). The
   GC UVLF in the case of a Schechter-like stellar mass function is itself
   Schechter-like, with a faint-end slope that is either the same as the
   underlying Schechter function ($\alpha$) or, if the input Schechter function
   slope is shallower than $-1.7$, is $\approx -1.7$ (see discussion below
   Eq.~\eqref{eq:4}).
 \label{fig:schech_vs_lognorm}
}
\end{figure}
The results of this paper generally assume that the observed log-normal
luminosity function of GCs is a manifestation of the initial
conditions. However, this may not be the case: an initially Schechter-like
luminosity function can evolve to a log-normal one over a Hubble time.

The Schechter \citeyearpar{schechter1976} luminosity function is parametrized by
a normalization ($\phi^*$), a characteristic luminosity ($L^*$), and a faint-end
slope ($\alpha$):
\begin{equation}
  \label{eq:schechter}
  \phi(L)=\frac{\phi^*}{L^*} \left(\frac{L}{L^*}\right)^{\alpha}\,\exp\left(-\frac{L}{L^*}\right)
\end{equation}
Analogously to the case of the log-normal luminosity function, we can consider
$L^*$ to be a function of time and compute the GC UVLF as an integral:
\begin{equation}
  \label{eq:2}
  \phi(L,t)\propto \int \phi(L, L^*(t)) \frac{dt}{\Delta t}
\end{equation}
(where I have assumed $\dot{n}_{\rm GC}$ is constant, consistent with the rest
of this paper; it is straightforward to generalize to a power-law dependence on
time). With the time evolution of $L^*$ given by
\begin{equation}
  \label{eq:1}
  L^*(t)=L_0\left(\frac{t}{t_0}\right)^{-b}\,,
\end{equation}
the luminosity function becomes
\begin{equation}
  \label{eq:3}
  \phi(L)=\frac{\phi^*}{L_0}\left(
    \frac{L}{L_0}\right)^{-1-1/b}\frac{t_0}{\Delta t} b^{-1} \,
  \left[ \gamma(\lambda, u)-\gamma(\lambda, u_0)\right]\,,
\end{equation}
where 
\begin{flalign}
  \label{eq:4}
\lambda&=\alpha+\frac{1}{b}+1\,,\\
u(t)&=u(L, t)=\frac{L}{L_0}\left(\frac{t}{t_0} \right)^b\,
\end{flalign}
and $\gamma(a,x)$ is the lower incomplete gamma function
\citep{abramowitz1972}. As with the log-normal case, the GCLF is sum over
individual components corresponding to the piecewise time ranges in
Eq.~\eqref{eq:t_var}.

The faint-end slope of $\phi(L)$ in the limit $L \ll L_0$ will be approximately
$-1-1/b$ if $\alpha > -1-1/b$ and $\alpha$ otherwise, as
$\gamma(\lambda,u) \propto u^{\lambda}$ as $u \rightarrow 0$. Notably, this
faint-end slope is identical to the log-normal case for relatively shallow
values of $\alpha$, which is a direct result of the time evolution of $L^*(t)$
-- determined by stellar evolution -- in Eq.~\eqref{eq:1}. If the faint-end
slope of the initial GC stellar mass function is steeper than $-(1+1/b)$, then
it is this initial faint-end slope that sets the overall slope of
$\phi_{\rm GC}(L)$.

Fig.~\ref{fig:schech_vs_lognorm} compares the GC UVLFs (after 1000
Myr of GC formation) for initial GC stellar mass functions that are
Schechter-like (with faint-end slope $-1.9$; black line) and log-normal
(gray). The input initial GC stellar mass functions are normalized such that the
number of GCs brighter than $\overline{M}$ (the peak value in the log-normal
stellar mass function) are equal. The resulting GC UVLFs are nearly identical
for bright magnitudes, while the Schechter-like initial GC stellar mass function
results in a higher amplitude for $\phi_{\rm GC}(L)$ at fainter magnitudes.

\label{lastpage}
\end{document}

%% file: extra_preamble.tex
\newcommand{\mstar}{M_{\star}}
\newcommand{\msun}{M_{\odot}}

\newcommand{\mpc}{{\rm Mpc}}

\newcommand{\muv}{{M_{\rm 1500}}}
\newcommand{\hst}{\textit{HST}}
\newcommand{\jwst}{\textit{JWST}}

\newcommand{\fesc}{f_{\rm esc}}
\newcommand{\mhalo}{M_{\rm halo}}

%% file: draft_clean.bbl
\begin{thebibliography}{}
\makeatletter
\relax
\def\mn@urlcharsother{\let\do\@makeother \do\$\do\&\do\#\do\^\do\_\do\%\do\~}
\def\mn@doi{\begingroup\mn@urlcharsother \@ifnextchar [ {\mn@doi@}
  {\mn@doi@[]}}
\def\mn@doi@[#1]#2{\def\@tempa{#1}\ifx\@tempa\@empty \href
  {http://dx.doi.org/#2} {doi:#2}\else \href {http://dx.doi.org/#2} {#1}\fi
  \endgroup}
\def\mn@eprint#1#2{\mn@eprint@#1:#2::\@nil}
\def\mn@eprint@arXiv#1{\href {http://arxiv.org/abs/#1} {{\tt arXiv:#1}}}
\def\mn@eprint@dblp#1{\href {http://dblp.uni-trier.de/rec/bibtex/#1.xml}
  {dblp:#1}}
\def\mn@eprint@#1:#2:#3:#4\@nil{\def\@tempa {#1}\def\@tempb {#2}\def\@tempc
  {#3}\ifx \@tempc \@empty \let \@tempc \@tempb \let \@tempb \@tempa \fi \ifx
  \@tempb \@empty \def\@tempb {arXiv}\fi \@ifundefined
  {mn@eprint@\@tempb}{\@tempb:\@tempc}{\expandafter \expandafter \csname
  mn@eprint@\@tempb\endcsname \expandafter{\@tempc}}}

\bibitem[\protect\citeauthoryear{{Abramowitz} \& {Stegun}}{{Abramowitz} \&
  {Stegun}}{1972}]{abramowitz1972}
{Abramowitz} M.,  {Stegun} I.~A.,  1972, {Handbook of Mathematical Functions}

\bibitem[\protect\citeauthoryear{{Anderson}, {Governato}, {Karcher}, {Quinn}
  \& {Wadsley}}{{Anderson} et~al.}{2017}]{anderson2017}
{Anderson} L.,  {Governato} F.,  {Karcher} M.,  {Quinn} T.,   {Wadsley} J.,
  2017, \mn@doi [\mnras] {10.1093/mnras/stx709}, \href
  {http://adsabs.harvard.edu/abs/2017MNRAS.468.4077A} {468, 4077}

\bibitem[\protect\citeauthoryear{{Bastian} \& {Lardo}}{{Bastian} \&
  {Lardo}}{2015}]{bastian2015}
{Bastian} N.,  {Lardo} C.,  2015, \mn@doi [\mnras] {10.1093/mnras/stv1661},
  \href {http://adsabs.harvard.edu/abs/2015MNRAS.453..357B} {453, 357}

\bibitem[\protect\citeauthoryear{{Bastian}, {Gieles}, {Lamers}, {Scheepmaker}
  \& {de Grijs}}{{Bastian} et~al.}{2005}]{bastian2005}
{Bastian} N.,  {Gieles} M.,  {Lamers} H.~J.~G.~L.~M.,  {Scheepmaker} R.~A.,
  {de Grijs} R.,  2005, \mn@doi [\aap] {10.1051/0004-6361:20041078}, \href
  {http://adsabs.harvard.edu/abs/2005A%26A...431..905B} {431, 905}

\bibitem[\protect\citeauthoryear{{Bastian}, {Hollyhead}  \&
  {Cabrera-Ziri}}{{Bastian} et~al.}{2014}]{bastian2014}
{Bastian} N.,  {Hollyhead} K.,   {Cabrera-Ziri} I.,  2014, \mn@doi [\mnras]
  {10.1093/mnras/stu1775}, \href
  {http://adsabs.harvard.edu/abs/2014MNRAS.445..378B} {445, 378}

\bibitem[\protect\citeauthoryear{{Basu}, {Gil}  \& {Auddy}}{{Basu}
  et~al.}{2015}]{basu2015}
{Basu} S.,  {Gil} M.,   {Auddy} S.,  2015, \mn@doi [\mnras]
  {10.1093/mnras/stv445}, \href
  {http://adsabs.harvard.edu/abs/2015MNRAS.449.2413B} {449, 2413}

\bibitem[\protect\citeauthoryear{{Baumgardt}, {Kroupa}  \&
  {Parmentier}}{{Baumgardt} et~al.}{2008}]{baumgardt2008}
{Baumgardt} H.,  {Kroupa} P.,   {Parmentier} G.,  2008, \mn@doi [\mnras]
  {10.1111/j.1365-2966.2007.12811.x}, \href
  {http://adsabs.harvard.edu/abs/2008MNRAS.384.1231B} {384, 1231}

\bibitem[\protect\citeauthoryear{{Binney} \& {Tremaine}}{{Binney} \&
  {Tremaine}}{1987}]{binney1987}
{Binney} J.,  {Tremaine} S.,  1987, {Galactic Dynamics}.
Princeton, NJ, Princeton University Press

\bibitem[\protect\citeauthoryear{{Bland-Hawthorn} \&
  {Freeman}}{{Bland-Hawthorn} \& {Freeman}}{2000}]{bland-hawthorn2000}
{Bland-Hawthorn} J.,  {Freeman} K.,  2000, \mn@doi [Science]
  {10.1126/science.287.5450.79}, \href
  {http://adsabs.harvard.edu/abs/2000Sci...287...79B} {287, 79}

\bibitem[\protect\citeauthoryear{{Bouwens}}{{Bouwens}}{2016}]{bouwens2016}
{Bouwens} R.,  2016, in {Mesinger} A.,  ed.,  Astrophysics and Space Science
  Library Vol. 423, Understanding the Epoch of Cosmic Reionization: Challenges
  and Progress. p.~111 (\mn@eprint {arXiv} {1511.01133}),
  \mn@doi{10.1007/978-3-319-21957-8_4}

\bibitem[\protect\citeauthoryear{{Bouwens}, {Illingworth}, {Oesch}, {Atek},
  {Lam}  \& {Stefanon}}{{Bouwens} et~al.}{2017}]{bouwens2017}
{Bouwens} R.~J.,  {Illingworth} G.~D.,  {Oesch} P.~A.,  {Atek} H.,  {Lam} D.,
  {Stefanon} M.,  2017, \mn@doi [\apj] {10.3847/1538-4357/aa74e4}, \href
  {http://adsabs.harvard.edu/abs/2017ApJ...843...41B} {843, 41}

\bibitem[\protect\citeauthoryear{{Boylan-Kolchin}}{{Boylan-Kolchin}}{2017}]{boylan-kolchin2017}
{Boylan-Kolchin} M.,  2017, \mn@doi [\mnras] {10.1093/mnras/stx2164}, \href
  {http://adsabs.harvard.edu/abs/2017MNRAS.472.3120B} {472, 3120}

\bibitem[\protect\citeauthoryear{{Boylan-Kolchin}, {Weisz}, {Johnson},
  {Bullock}, {Conroy}  \& {Fitts}}{{Boylan-Kolchin}
  et~al.}{2015}]{boylan-kolchin2015}
{Boylan-Kolchin} M.,  {Weisz} D.~R.,  {Johnson} B.~D.,  {Bullock} J.~S.,
  {Conroy} C.,   {Fitts} A.,  2015, \mn@doi [\mnras] {10.1093/mnras/stv1736},
  \href {http://adsabs.harvard.edu/abs/2015MNRAS.453.1503B} {453, 1503}

\bibitem[\protect\citeauthoryear{{Brodie} \& {Strader}}{{Brodie} \&
  {Strader}}{2006}]{brodie2006}
{Brodie} J.~P.,  {Strader} J.,  2006, \mn@doi [\araa]
  {10.1146/annurev.astro.44.051905.092441}, \href
  {http://adsabs.harvard.edu/abs/2006ARA%26A..44..193B} {44, 193}

\bibitem[\protect\citeauthoryear{{Bromm} \& {Yoshida}}{{Bromm} \&
  {Yoshida}}{2011}]{bromm2011}
{Bromm} V.,  {Yoshida} N.,  2011, \mn@doi [\araa]
  {10.1146/annurev-astro-081710-102608}, \href
  {http://adsabs.harvard.edu/abs/2011ARA%26A..49..373B} {49, 373}

\bibitem[\protect\citeauthoryear{{Carlberg}}{{Carlberg}}{2002}]{carlberg2002}
{Carlberg} R.~G.,  2002, \mn@doi [\apj] {10.1086/340500}, \href
  {http://adsabs.harvard.edu/abs/2002ApJ...573...60C} {573, 60}

\bibitem[\protect\citeauthoryear{{Chandrasekhar}}{{Chandrasekhar}}{1942}]{chandrasekhar1942}
{Chandrasekhar} S.,  1942, {Principles of stellar dynamics}

\bibitem[\protect\citeauthoryear{{Charbonnel}}{{Charbonnel}}{2016}]{charbonnel2016a}
{Charbonnel} C.,  2016, in {Moraux} E.,  {Lebreton} Y.,   {Charbonnel} C.,
  eds,  EAS Publications Series Vol. 80, EAS Publications Series. pp 177--226
  (\mn@eprint {arXiv} {1611.08855}), \mn@doi{10.1051/eas/1680006}

\bibitem[\protect\citeauthoryear{{Conroy} \& {Spergel}}{{Conroy} \&
  {Spergel}}{2011}]{conroy2011}
{Conroy} C.,  {Spergel} D.~N.,  2011, \mn@doi [\apj]
  {10.1088/0004-637X/726/1/36}, \href
  {http://adsabs.harvard.edu/abs/2011ApJ...726...36C} {726, 36}

\bibitem[\protect\citeauthoryear{{D'Ercole}, {Vesperini}, {D'Antona},
  {McMillan}  \& {Recchi}}{{D'Ercole} et~al.}{2008}]{dercole2008}
{D'Ercole} A.,  {Vesperini} E.,  {D'Antona} F.,  {McMillan} S.~L.~W.,
  {Recchi} S.,  2008, \mn@doi [\mnras] {10.1111/j.1365-2966.2008.13915.x},
  \href {http://adsabs.harvard.edu/abs/2008MNRAS.391..825D} {391, 825}

\bibitem[\protect\citeauthoryear{{Decressin}, {Meynet}, {Charbonnel},
  {Prantzos}  \& {Ekstr{\"o}m}}{{Decressin} et~al.}{2007}]{decressin2007}
{Decressin} T.,  {Meynet} G.,  {Charbonnel} C.,  {Prantzos} N.,   {Ekstr{\"o}m}
  S.,  2007, \mn@doi [\aap] {10.1051/0004-6361:20066013}, \href
  {http://adsabs.harvard.edu/abs/2007A%26A...464.1029D} {464, 1029}

\bibitem[\protect\citeauthoryear{{Denissenkov} \& {Hartwick}}{{Denissenkov} \&
  {Hartwick}}{2014}]{denissenkov2014}
{Denissenkov} P.~A.,  {Hartwick} F.~D.~A.,  2014, \mn@doi [\mnras]
  {10.1093/mnrasl/slt133}, \href
  {http://adsabs.harvard.edu/abs/2014MNRAS.437L..21D} {437, L21}

\bibitem[\protect\citeauthoryear{{Eggen}, {Lynden-Bell}  \& {Sandage}}{{Eggen}
  et~al.}{1962}]{eggen1962}
{Eggen} O.~J.,  {Lynden-Bell} D.,   {Sandage} A.~R.,  1962, \mn@doi [\apj]
  {10.1086/147433}, \href {http://adsabs.harvard.edu/abs/1962ApJ...136..748E}
  {136, 748}

\bibitem[\protect\citeauthoryear{{Fall} \& {Rees}}{{Fall} \&
  {Rees}}{1977}]{fall1977}
{Fall} S.~M.,  {Rees} M.~J.,  1977, \mn@doi [\mnras] {10.1093/mnras/181.1.37P},
  \href {http://adsabs.harvard.edu/abs/1977MNRAS.181P..37F} {181, 37P}

\bibitem[\protect\citeauthoryear{{Fall} \& {Rees}}{{Fall} \&
  {Rees}}{1985}]{fall1985}
{Fall} S.~M.,  {Rees} M.~J.,  1985, \mn@doi [\apj] {10.1086/163585}, \href
  {http://adsabs.harvard.edu/abs/1985ApJ...298...18F} {298, 18}

\bibitem[\protect\citeauthoryear{{Fall} \& {Zhang}}{{Fall} \&
  {Zhang}}{2001}]{fall2001}
{Fall} S.~M.,  {Zhang} Q.,  2001, \mn@doi [\apj] {10.1086/323358}, \href
  {http://adsabs.harvard.edu/abs/2001ApJ...561..751F} {561, 751}

\bibitem[\protect\citeauthoryear{{Fall}, {Chandar}  \& {Whitmore}}{{Fall}
  et~al.}{2005}]{fall2005}
{Fall} S.~M.,  {Chandar} R.,   {Whitmore} B.~C.,  2005, \mn@doi [\apjl]
  {10.1086/496878}, \href {http://adsabs.harvard.edu/abs/2005ApJ...631L.133F}
  {631, L133}

\bibitem[\protect\citeauthoryear{{Finkelstein}}{{Finkelstein}}{2016}]{finkelstein2016}
{Finkelstein} S.~L.,  2016, \mn@doi [\pasa] {10.1017/pasa.2016.26}, \href
  {http://adsabs.harvard.edu/abs/2016PASA...33...37F} {33, e037}

\bibitem[\protect\citeauthoryear{{Finkelstein} et~al.,}{{Finkelstein}
  et~al.}{2015}]{finkelstein2015}
{Finkelstein} S.~L.,  et~al., 2015, \mn@doi [\apj]
  {10.1088/0004-637X/810/1/71}, \href
  {http://adsabs.harvard.edu/abs/2015ApJ...810...71F} {810, 71}

\bibitem[\protect\citeauthoryear{{Forbes}}{{Forbes}}{2017}]{forbes2017}
{Forbes} D.~A.,  2017, \mn@doi [\mnras] {10.1093/mnrasl/slx148}, \href
  {http://adsabs.harvard.edu/abs/2017MNRAS.472L.104F} {472, L104}

\bibitem[\protect\citeauthoryear{{Gnedin}}{{Gnedin}}{2016}]{gnedin2016}
{Gnedin} N.~Y.,  2016, \mn@doi [\apjl] {10.3847/2041-8205/825/2/L17}, \href
  {http://adsabs.harvard.edu/abs/2016ApJ...825L..17G} {825, L17}

\bibitem[\protect\citeauthoryear{{Gnedin}, {Hernquist}  \& {Ostriker}}{{Gnedin}
  et~al.}{1999}]{gnedin1999}
{Gnedin} O.~Y.,  {Hernquist} L.,   {Ostriker} J.~P.,  1999, \mn@doi [\apj]
  {10.1086/306910}, \href
  {http://adsabs.harvard.edu/cgi-bin/nph-bib_query?bibcode=1999ApJ...514..109G&db_key=AST}
  {514, 109}

\bibitem[\protect\citeauthoryear{{Gratton}, {Carretta}  \&
  {Bragaglia}}{{Gratton} et~al.}{2012}]{gratton2012}
{Gratton} R.~G.,  {Carretta} E.,   {Bragaglia} A.,  2012, \mn@doi [\aapr]
  {10.1007/s00159-012-0050-3}, \href
  {http://adsabs.harvard.edu/abs/2012A%26ARv..20...50G} {20, 50}

\bibitem[\protect\citeauthoryear{{Harris}}{{Harris}}{1991}]{harris1991}
{Harris} W.~E.,  1991, \mn@doi [\araa] {10.1146/annurev.aa.29.090191.002551},
  \href {http://adsabs.harvard.edu/abs/1991ARA%26A..29..543H} {29, 543}

\bibitem[\protect\citeauthoryear{{Harris}}{{Harris}}{1996}]{harris1996}
{Harris} W.~E.,  1996, \mn@doi [\aj] {10.1086/118116}, \href
  {http://adsabs.harvard.edu/abs/1996AJ....112.1487H} {112, 1487}

\bibitem[\protect\citeauthoryear{{Hollyhead}, {Bastian}, {Adamo},
  {Silva-Villa}, {Dale}, {Ryon}  \& {Gazak}}{{Hollyhead}
  et~al.}{2015}]{hollyhead2015}
{Hollyhead} K.,  {Bastian} N.,  {Adamo} A.,  {Silva-Villa} E.,  {Dale} J.,
  {Ryon} J.~E.,   {Gazak} Z.,  2015, \mn@doi [\mnras] {10.1093/mnras/stv331},
  \href {http://adsabs.harvard.edu/abs/2015MNRAS.449.1106H} {449, 1106}

\bibitem[\protect\citeauthoryear{{Howard}, {Pudritz}, {Harris}  \&
  {Klessen}}{{Howard} et~al.}{2017}]{howard2017a}
{Howard} C.~S.,  {Pudritz} R.~E.,  {Harris} B.~E.,   {Klessen} R.~S.,  2017,
  {arXiv:1710.04283 [astro-ph]}, \href
  {http://adsabs.harvard.edu/abs/2017arXiv171004283H} {}

\bibitem[\protect\citeauthoryear{{Hudson} \& {Robison}}{{Hudson} \&
  {Robison}}{2017}]{hudson2017}
{Hudson} M.~J.,  {Robison} B.,  2017, {arXiv:1707.02609 [astro-ph]}, \href
  {http://adsabs.harvard.edu/abs/2017arXiv170702609H} {}

\bibitem[\protect\citeauthoryear{Hunter}{Hunter}{2007}]{matplotlib}
Hunter J.~D.,  2007, Computing In Science \& Engineering, 9, 90

\bibitem[\protect\citeauthoryear{{Ishigaki}, {Kawamata}, {Ouchi}, {Oguri},
  {Shimasaku}  \& {Ono}}{{Ishigaki} et~al.}{2017}]{ishigaki2017}
{Ishigaki} M.,  {Kawamata} R.,  {Ouchi} M.,  {Oguri} M.,  {Shimasaku} K.,
  {Ono} Y.,  2017, {arXiv:1702.04867 [astro-ph]}, \href
  {http://adsabs.harvard.edu/abs/2017arXiv170204867I} {}

\bibitem[\protect\citeauthoryear{{Katz} \& {Ricotti}}{{Katz} \&
  {Ricotti}}{2013}]{katz2013}
{Katz} H.,  {Ricotti} M.,  2013, \mn@doi [\mnras] {10.1093/mnras/stt676}, \href
  {http://adsabs.harvard.edu/abs/2013MNRAS.432.3250K} {432, 3250}

\bibitem[\protect\citeauthoryear{{Katz} \& {Ricotti}}{{Katz} \&
  {Ricotti}}{2014}]{katz2014}
{Katz} H.,  {Ricotti} M.,  2014, \mn@doi [\mnras] {10.1093/mnras/stu1489},
  \href {http://adsabs.harvard.edu/abs/2014MNRAS.444.2377K} {444, 2377}

\bibitem[\protect\citeauthoryear{{Kawamata}, {Ishigaki}, {Shimasaku}, {Oguri}
  \& {Ouchi}}{{Kawamata} et~al.}{2015}]{kawamata2015}
{Kawamata} R.,  {Ishigaki} M.,  {Shimasaku} K.,  {Oguri} M.,   {Ouchi} M.,
  2015, \mn@doi [\apj] {10.1088/0004-637X/804/2/103}, \href
  {http://adsabs.harvard.edu/abs/2015ApJ...804..103K} {804, 103}

\bibitem[\protect\citeauthoryear{{Kawamata}, {Ishigaki}, {Shimasaku}, {Oguri},
  {Ouchi}  \& {Tanigawa}}{{Kawamata} et~al.}{2018}]{kawamata2018}
{Kawamata} R.,  {Ishigaki} M.,  {Shimasaku} K.,  {Oguri} M.,  {Ouchi} M.,
  {Tanigawa} S.,  2018, \mn@doi [\apj] {10.3847/1538-4357/aaa6cf}, \href
  {http://adsabs.harvard.edu/abs/2018ApJ...855....4K} {855, 4}

\bibitem[\protect\citeauthoryear{{Kim} et~al.,}{{Kim} et~al.}{2018}]{kim2018}
{Kim} J.-h.,  et~al., 2018, \mn@doi [\mnras] {10.1093/mnras/stx2994}, \href
  {http://adsabs.harvard.edu/abs/2018MNRAS.474.4232K} {474, 4232}

\bibitem[\protect\citeauthoryear{{Kimm}, {Cen}, {Rosdahl}  \& {Yi}}{{Kimm}
  et~al.}{2016}]{kimm2016}
{Kimm} T.,  {Cen} R.,  {Rosdahl} J.,   {Yi} S.~K.,  2016, \mn@doi [\apj]
  {10.3847/0004-637X/823/1/52}, \href
  {http://adsabs.harvard.edu/abs/2016ApJ...823...52K} {823, 52}

\bibitem[\protect\citeauthoryear{{Kimm}, {Katz}, {Haehnelt}, {Rosdahl},
  {Devriendt}  \& {Slyz}}{{Kimm} et~al.}{2017}]{kimm2017}
{Kimm} T.,  {Katz} H.,  {Haehnelt} M.,  {Rosdahl} J.,  {Devriendt} J.,   {Slyz}
  A.,  2017, \mn@doi [\mnras] {10.1093/mnras/stx052}, \href
  {http://adsabs.harvard.edu/abs/2017MNRAS.466.4826K} {466, 4826}

\bibitem[\protect\citeauthoryear{{Kroupa}}{{Kroupa}}{2001}]{kroupa2001}
{Kroupa} P.,  2001, \mn@doi [\mnras] {10.1046/j.1365-8711.2001.04022.x}, \href
  {http://adsabs.harvard.edu/abs/2001MNRAS.322..231K} {322, 231}

\bibitem[\protect\citeauthoryear{{Krumholz}}{{Krumholz}}{2015}]{krumholz2015}
{Krumholz} M.~R.,  2015, {arXiv:1511.03457 [astro-ph]}, \href
  {http://adsabs.harvard.edu/abs/2015arXiv151103457K} {}

\bibitem[\protect\citeauthoryear{{Kuhlen} \& {Faucher-Gigu{\`e}re}}{{Kuhlen} \&
  {Faucher-Gigu{\`e}re}}{2012}]{kuhlen2012a}
{Kuhlen} M.,  {Faucher-Gigu{\`e}re} C.-A.,  2012, \mn@doi [\mnras]
  {10.1111/j.1365-2966.2012.20924.x}, \href
  {http://adsabs.harvard.edu/abs/2012MNRAS.423..862K} {423, 862}

\bibitem[\protect\citeauthoryear{{Larsen}}{{Larsen}}{2002}]{larsen2002}
{Larsen} S.~S.,  2002, \mn@doi [\aj] {10.1086/342381}, \href
  {http://adsabs.harvard.edu/abs/2002AJ....124.1393L} {124, 1393}

\bibitem[\protect\citeauthoryear{{Larson}}{{Larson}}{1990}]{larson1990}
{Larson} R.~B.,  1990, \mn@doi [\pasp] {10.1086/132694}, \href
  {http://adsabs.harvard.edu/abs/1990PASP..102..709L} {102, 709}

\bibitem[\protect\citeauthoryear{{Ma}, {Kasen}, {Hopkins},
  {Faucher-Gigu{\`e}re}, {Quataert}, {Kere{\v s}}  \& {Murray}}{{Ma}
  et~al.}{2015}]{ma2015}
{Ma} X.,  {Kasen} D.,  {Hopkins} P.~F.,  {Faucher-Gigu{\`e}re} C.-A.,
  {Quataert} E.,  {Kere{\v s}} D.,   {Murray} N.,  2015, \mn@doi [\mnras]
  {10.1093/mnras/stv1679}, \href
  {http://adsabs.harvard.edu/abs/2015MNRAS.453..960M} {453, 960}

\bibitem[\protect\citeauthoryear{{Ma} et~al.,}{{Ma} et~al.}{2017}]{ma2017a}
{Ma} X.,  et~al., 2017, {arXiv:1710.00008 [astro-ph]}, \href
  {http://adsabs.harvard.edu/abs/2017arXiv171000008M} {}

\bibitem[\protect\citeauthoryear{{Madau}}{{Madau}}{2017}]{madau2017}
{Madau} P.,  2017, \mn@doi [\apj] {10.3847/1538-4357/aa9715}, \href
  {http://adsabs.harvard.edu/abs/2017ApJ...851...50M} {851, 50}

\bibitem[\protect\citeauthoryear{{Mar{\'{\i}}n-Franch}
  et~al.,}{{Mar{\'{\i}}n-Franch} et~al.}{2009}]{marin-franch2009}
{Mar{\'{\i}}n-Franch} A.,  et~al., 2009, \mn@doi [\apj]
  {10.1088/0004-637X/694/2/1498}, \href
  {http://adsabs.harvard.edu/abs/2009ApJ...694.1498M} {694, 1498}

\bibitem[\protect\citeauthoryear{{McLure} et~al.,}{{McLure}
  et~al.}{2013}]{mclure2013}
{McLure} R.~J.,  et~al., 2013, \mn@doi [\mnras] {10.1093/mnras/stt627}, \href
  {http://adsabs.harvard.edu/abs/2013MNRAS.432.2696M} {432, 2696}

\bibitem[\protect\citeauthoryear{{Murali} \& {Weinberg}}{{Murali} \&
  {Weinberg}}{1997}]{murali1997}
{Murali} C.,  {Weinberg} M.~D.,  1997, \mn@doi [\mnras]
  {10.1093/mnras/291.4.717}, \href
  {http://adsabs.harvard.edu/abs/1997MNRAS.291..717M} {291, 717}

\bibitem[\protect\citeauthoryear{{Oliphant}}{{Oliphant}}{2007}]{scipy}
{Oliphant} T.~E.,  2007, \mn@doi [Computing in Science Engineering]
  {10.1109/MCSE.2007.58}, 9, 10

\bibitem[\protect\citeauthoryear{{Ostriker}, {Spitzer}  \&
  {Chevalier}}{{Ostriker} et~al.}{1972}]{ostriker1972}
{Ostriker} J.~P.,  {Spitzer} L.~J.,   {Chevalier} R.~A.,  1972, \apjl, \href
  {http://adsabs.harvard.edu/cgi-bin/nph-bib_query?bibcode=1972ApJ...176L..51O&db_key=AST}
  {176, L51}

\bibitem[\protect\citeauthoryear{{Paardekooper}, {Khochfar}  \& {Dalla
  Vecchia}}{{Paardekooper} et~al.}{2015}]{paardekooper2015}
{Paardekooper} J.-P.,  {Khochfar} S.,   {Dalla Vecchia} C.,  2015, \mn@doi
  [\mnras] {10.1093/mnras/stv1114}, \href
  {http://adsabs.harvard.edu/abs/2015MNRAS.451.2544P} {451, 2544}

\bibitem[\protect\citeauthoryear{{Parmentier} \& {Gilmore}}{{Parmentier} \&
  {Gilmore}}{2007}]{parmentier2007}
{Parmentier} G.,  {Gilmore} G.,  2007, \mn@doi [\mnras]
  {10.1111/j.1365-2966.2007.11611.x}, \href
  {http://adsabs.harvard.edu/abs/2007MNRAS.377..352P} {377, 352}

\bibitem[\protect\citeauthoryear{{Peebles} \& {Dicke}}{{Peebles} \&
  {Dicke}}{1968}]{peebles1968}
{Peebles} P.~J.~E.,  {Dicke} R.~H.,  1968, \mn@doi [\apj] {10.1086/149811},
  \href {http://adsabs.harvard.edu/abs/1968ApJ...154..891P} {154, 891}

\bibitem[\protect\citeauthoryear{P\'erez \& Granger}{P\'erez \&
  Granger}{2007}]{ipython}
P\'erez F.,  Granger B.~E.,  2007, \mn@doi [Computing in Science and
  Engineering] {10.1109/MCSE.2007.53}, 9, 21

\bibitem[\protect\citeauthoryear{{Planck Collaboration} et~al.,}{{Planck
  Collaboration} et~al.}{2016}]{planck2015}
{Planck Collaboration} et~al., 2016, \mn@doi [\aap]
  {10.1051/0004-6361/201525830}, \href
  {http://adsabs.harvard.edu/abs/2016A%26A...594A..13P} {594, A13}

\bibitem[\protect\citeauthoryear{{Portegies Zwart} \& {McMillan}}{{Portegies
  Zwart} \& {McMillan}}{2000}]{portegies-zwart2000}
{Portegies Zwart} S.~F.,  {McMillan} S.~L.~W.,  2000, \mn@doi [\apjl]
  {10.1086/312422}, \href {http://adsabs.harvard.edu/abs/2000ApJ...528L..17P}
  {528, L17}

\bibitem[\protect\citeauthoryear{{Portegies Zwart}, {McMillan}  \&
  {Gieles}}{{Portegies Zwart} et~al.}{2010}]{portegies-zwart2010}
{Portegies Zwart} S.~F.,  {McMillan} S.~L.~W.,   {Gieles} M.,  2010, \mn@doi
  [\araa] {10.1146/annurev-astro-081309-130834}, \href
  {http://adsabs.harvard.edu/abs/2010ARA%26A..48..431P} {48, 431}

\bibitem[\protect\citeauthoryear{{Prieto} \& {Gnedin}}{{Prieto} \&
  {Gnedin}}{2008}]{prieto2008}
{Prieto} J.~L.,  {Gnedin} O.~Y.,  2008, \mn@doi [\apj] {10.1086/591777}, \href
  {http://adsabs.harvard.edu/abs/2008ApJ...689..919P} {689, 919}

\bibitem[\protect\citeauthoryear{{Renzini}}{{Renzini}}{2017}]{renzini2017}
{Renzini} A.,  2017, \mn@doi [\mnras] {10.1093/mnrasl/slx057}, \href
  {http://adsabs.harvard.edu/abs/2017MNRAS.469L..63R} {469, L63}

\bibitem[\protect\citeauthoryear{{Renzini} et~al.,}{{Renzini}
  et~al.}{2015}]{renzini2015}
{Renzini} A.,  et~al., 2015, \mn@doi [\mnras] {10.1093/mnras/stv2268}, \href
  {http://adsabs.harvard.edu/abs/2015MNRAS.454.4197R} {454, 4197}

\bibitem[\protect\citeauthoryear{{Ricotti}}{{Ricotti}}{2002}]{ricotti2002}
{Ricotti} M.,  2002, \mn@doi [\mnras] {10.1046/j.1365-8711.2002.05990.x}, \href
  {http://adsabs.harvard.edu/abs/2002MNRAS.336L..33R} {336, L33}

\bibitem[\protect\citeauthoryear{{Robertson} et~al.,}{{Robertson}
  et~al.}{2013}]{robertson2013}
{Robertson} B.~E.,  et~al., 2013, \mn@doi [\apj] {10.1088/0004-637X/768/1/71},
  \href {http://adsabs.harvard.edu/abs/2013ApJ...768...71R} {768, 71}

\bibitem[\protect\citeauthoryear{{Rosolowsky}}{{Rosolowsky}}{2005}]{rosolowsky2005}
{Rosolowsky} E.,  2005, \mn@doi [\pasp] {10.1086/497582}, \href
  {http://esoads.eso.org/abs/2005PASP..117.1403R} {117, 1403}

\bibitem[\protect\citeauthoryear{{Schaerer} \& {Charbonnel}}{{Schaerer} \&
  {Charbonnel}}{2011}]{schaerer2011}
{Schaerer} D.,  {Charbonnel} C.,  2011, \mn@doi [\mnras]
  {10.1111/j.1365-2966.2011.18304.x}, \href
  {http://adsabs.harvard.edu/abs/2011MNRAS.413.2297S} {413, 2297}

\bibitem[\protect\citeauthoryear{{Schechter}}{{Schechter}}{1976}]{schechter1976}
{Schechter} P.,  1976, \mn@doi [\apj] {10.1086/154079}, \href
  {http://adsabs.harvard.edu/abs/1976ApJ...203..297S} {203, 297}

\bibitem[\protect\citeauthoryear{{Schenker} et~al.,}{{Schenker}
  et~al.}{2013}]{schenker2013}
{Schenker} M.~A.,  et~al., 2013, \mn@doi [\apj] {10.1088/0004-637X/768/2/196},
  \href {http://adsabs.harvard.edu/abs/2013ApJ...768..196S} {768, 196}

\bibitem[\protect\citeauthoryear{{Searle} \& {Zinn}}{{Searle} \&
  {Zinn}}{1978}]{searle1978}
{Searle} L.,  {Zinn} R.,  1978, \mn@doi [\apj] {10.1086/156499}, \href
  {http://adsabs.harvard.edu/abs/1978ApJ...225..357S} {225, 357}

\bibitem[\protect\citeauthoryear{{Sheth}, {Mo}  \& {Tormen}}{{Sheth}
  et~al.}{2001}]{sheth2001}
{Sheth} R.~K.,  {Mo} H.~J.,   {Tormen} G.,  2001, \mn@doi [\mnras]
  {10.1046/j.1365-8711.2001.04006.x}, \href
  {http://adsabs.harvard.edu/abs/2001MNRAS.323....1S} {323, 1}

\bibitem[\protect\citeauthoryear{{Shull}, {Harness}, {Trenti}  \&
  {Smith}}{{Shull} et~al.}{2012}]{shull2012}
{Shull} J.~M.,  {Harness} A.,  {Trenti} M.,   {Smith} B.~D.,  2012, \mn@doi
  [\apj] {10.1088/0004-637X/747/2/100}, \href
  {http://adsabs.harvard.edu/abs/2012ApJ...747..100S} {747, 100}

\bibitem[\protect\citeauthoryear{{Solomon}, {Rivolo}, {Barrett}  \&
  {Yahil}}{{Solomon} et~al.}{1987}]{solomon1987}
{Solomon} P.~M.,  {Rivolo} A.~R.,  {Barrett} J.,   {Yahil} A.,  1987, \mn@doi
  [\apj] {10.1086/165493}, \href
  {http://esoads.eso.org/abs/1987ApJ...319..730S} {319, 730}

\bibitem[\protect\citeauthoryear{{Spitzer}}{{Spitzer}}{1958}]{spitzer1958}
{Spitzer} Jr. L.,  1958, \mn@doi [\apj] {10.1086/146435}, \href
  {http://adsabs.harvard.edu/abs/1958ApJ...127...17S} {127, 17}

\bibitem[\protect\citeauthoryear{{Spitzer}}{{Spitzer}}{1987}]{spitzer1987}
{Spitzer} L.,  1987, {Dynamical Evolution of Globular Clusters}.
Princeton, NJ, Princeton University Press

\bibitem[\protect\citeauthoryear{{Stark}}{{Stark}}{2016}]{stark2016}
{Stark} D.~P.,  2016, \mn@doi [\araa] {10.1146/annurev-astro-081915-023417},
  \href {http://adsabs.harvard.edu/abs/2016ARA%26A..54..761S} {54, 761}

\bibitem[\protect\citeauthoryear{{Trebitsch}, {Blaizot}, {Rosdahl}, {Devriendt}
   \& {Slyz}}{{Trebitsch} et~al.}{2017}]{trebitsch2017}
{Trebitsch} M.,  {Blaizot} J.,  {Rosdahl} J.,  {Devriendt} J.,   {Slyz} A.,
  2017, \mn@doi [\mnras] {10.1093/mnras/stx1060}, \href
  {http://adsabs.harvard.edu/abs/2017MNRAS.470..224T} {470, 224}

\bibitem[\protect\citeauthoryear{{Van Der Walt}, {Colbert}  \&
  {Varoquaux}}{{Van Der Walt} et~al.}{2011}]{numpy}
{Van Der Walt} S.,  {Colbert} S.~C.,   {Varoquaux} G.,  2011, {arXiv:1102.1523
  [astro-ph]}, \href {http://adsabs.harvard.edu/abs/2011arXiv1102.1523V} {}

\bibitem[\protect\citeauthoryear{VandenBerg, Bolte  \& Stetson}{VandenBerg
  et~al.}{1996}]{vandenberg1996}
VandenBerg D.~A.,  Bolte M.,   Stetson P.~B.,  1996, \mn@doi [Annual Review of
  Astronomy and Astrophysics] {10.1146/annurev.astro.34.1.461}, 34, 461

\bibitem[\protect\citeauthoryear{{VandenBerg}, {Brogaard}, {Leaman}  \&
  {Casagrande}}{{VandenBerg} et~al.}{2013}]{vandenberg2013}
{VandenBerg} D.~A.,  {Brogaard} K.,  {Leaman} R.,   {Casagrande} L.,  2013,
  \mn@doi [\apj] {10.1088/0004-637X/775/2/134}, \href
  {http://adsabs.harvard.edu/abs/2013ApJ...775..134V} {775, 134}

\bibitem[\protect\citeauthoryear{{Vanzella} et~al.,}{{Vanzella}
  et~al.}{2017}]{vanzella2017}
{Vanzella} E.,  et~al., 2017, \mn@doi [\mnras] {10.1093/mnras/stx351}, \href
  {http://adsabs.harvard.edu/abs/2017MNRAS.467.4304V} {467, 4304}

\bibitem[\protect\citeauthoryear{{Villegas} et~al.,}{{Villegas}
  et~al.}{2010}]{villegas2010}
{Villegas} D.,  et~al., 2010, \mn@doi [\apj] {10.1088/0004-637X/717/2/603},
  \href {http://adsabs.harvard.edu/abs/2010ApJ...717..603V} {717, 603}

\bibitem[\protect\citeauthoryear{{Weisz} \& {Boylan-Kolchin}}{{Weisz} \&
  {Boylan-Kolchin}}{2017}]{weisz2017}
{Weisz} D.~R.,  {Boylan-Kolchin} M.,  2017, \mn@doi [\mnras]
  {10.1093/mnrasl/slx043}, 469, L83

\bibitem[\protect\citeauthoryear{{Whitmore} \& {Schweizer}}{{Whitmore} \&
  {Schweizer}}{1995}]{whitmore1995}
{Whitmore} B.~C.,  {Schweizer} F.,  1995, \mn@doi [\aj] {10.1086/117334}, \href
  {http://adsabs.harvard.edu/abs/1995AJ....109..960W} {109, 960}

\bibitem[\protect\citeauthoryear{{Wise}, {Demchenko}, {Halicek}, {Norman},
  {Turk}, {Abel}  \& {Smith}}{{Wise} et~al.}{2014}]{wise2014}
{Wise} J.~H.,  {Demchenko} V.~G.,  {Halicek} M.~T.,  {Norman} M.~L.,  {Turk}
  M.~J.,  {Abel} T.,   {Smith} B.~D.,  2014, \mn@doi [\mnras]
  {10.1093/mnras/stu979}, \href
  {http://adsabs.harvard.edu/abs/2014MNRAS.442.2560W} {442, 2560}

\bibitem[\protect\citeauthoryear{{Xu}, {Wise}, {Norman}, {Ahn}  \&
  {O'Shea}}{{Xu} et~al.}{2016}]{xu2016}
{Xu} H.,  {Wise} J.~H.,  {Norman} M.~L.,  {Ahn} K.,   {O'Shea} B.~W.,  2016,
  \mn@doi [\apj] {10.3847/1538-4357/833/1/84}, \href
  {http://adsabs.harvard.edu/abs/2016ApJ...833...84X} {833, 84}

\bibitem[\protect\citeauthoryear{{Zackrisson} et~al.,}{{Zackrisson}
  et~al.}{2017}]{zackrisson2017}
{Zackrisson} E.,  et~al., 2017, \mn@doi [\apj] {10.3847/1538-4357/836/1/78},
  \href {http://adsabs.harvard.edu/abs/2017ApJ...836...78Z} {836, 78}

\bibitem[\protect\citeauthoryear{{Zick}, {Weisz}  \& {Boylan-Kolchin}}{{Zick}
  et~al.}{2018}]{zick2018}
{Zick} T.~O.,  {Weisz} D.~R.,   {Boylan-Kolchin} M.,  2018, {arXiv:1802.06801
  [astro-ph]}, \href {http://adsabs.harvard.edu/abs/2018arXiv180206801Z} {}

\makeatother
\end{thebibliography}
